\documentclass[manuscript,screen]{acmart}

\usepackage[utf8]{inputenc}

\usepackage{lscape}
\usepackage{url}

\usepackage{graphicx}
\graphicspath{{images/}}

\usepackage{csquotes}

\usepackage{tcolorbox}

\usepackage{todonotes}

\usepackage{tabularx}
\usepackage{longtable}

\usepackage{soul}

\AtBeginDocument{%
  \providecommand\BibTeX{{%
    \normalfont B\kern-0.5em{\scshape i\kern-0.25em b}\kern-0.8em\TeX}}}

\acmConference[ACM TOSEM]{Submission}
\acmBooktitle{Accepted in ACM TOSEM}
\setcopyright{none}
\renewcommand\footnotetextcopyrightpermission[1]{}
\begin{document}

\title{Software Engineering for AI-Based Systems: A Survey}

\author{Silverio Martínez-Fernández}
\email{silverio.martinez@upc.edu}
\orcid{0000-0001-9928-133X}
\affiliation{%
  \institution{Universitat Politècnica de Catalunya - BarcelonaTech}
  \streetaddress{Jordi Girona, 1-3, omega oficina 004}
  \city{Barcelona}
  \state{Catalunya}
  \country{Spain}
  \postcode{08034}
}

\author{Justus Bogner}
\affiliation{
    \institution{University of Stuttgart, Institute of Software Engineering}
    \streetaddress{Universitätsstr. 38}
    \city{Stuttgart}
    \country{Germany}
    \postcode{70569}
}
\email{justus.bogner@iste.uni-stuttgart.de}

\author{Xavier Franch}
\affiliation{%
  \institution{Universitat Politècnica de Catalunya - BarcelonaTech}
  \streetaddress{Jordi Girona, 1-3, omega}
  \city{Barcelona}
  \state{Catalunya}
  \country{Spain}
}

\author{Marc Oriol}
\affiliation{%
  \institution{Universitat Politècnica de Catalunya - BarcelonaTech}
  \streetaddress{Jordi Girona, 1-3, omega}
  \city{Barcelona}
  \state{Catalunya}
  \country{Spain}}

\author{Julien Siebert}
\affiliation{%
  \institution{Fraunhofer Institute for Experimental Software Engineering IESE}
  \streetaddress{Fraunhofer-Platz 1}
  \city{Kaiserslautern}
  \country{Germany}
  \postcode{67663}}
\email{julien.siebert@iese.fraunhofer.de}

\author{Adam Trendowicz}
\affiliation{%
  \institution{Fraunhofer Institute for Experimental Software Engineering IESE}
  \streetaddress{Fraunhofer-Platz 1}
  \city{Kaiserslautern}
  \country{Germany}
  \postcode{67663}}
\email{adam.trendowicz@iese.fraunhofer.de}

\author{Anna Maria Vollmer}
\affiliation{%
  \institution{Fraunhofer Institute for Experimental Software Engineering IESE}
  \streetaddress{Fraunhofer-Platz 1}
  \city{Kaiserslautern}
  \country{Germany}
  \postcode{67663}}
\email{anna-maria.vollmer@iese.fraunhofer.de}

\author{Stefan Wagner}
\affiliation{
    \institution{University of Stuttgart, Institute of Software Engineering}
    \streetaddress{Universitätsstr. 38}
    \city{Stuttgart}
    \country{Germany}
    \postcode{70569}
}
\email{stefan.wagner@iste.uni-stuttgart.de}

\renewcommand{\shortauthors}{Martínez-Fernández, et al.}

\begin{abstract}
%For the abstract, there is a limit of 200 words!!
AI-based systems are software systems with functionalities enabled by at least one AI component (e.g., for image-, speech-recognition, and autonomous driving). AI-based systems are becoming pervasive in society due to advances in AI. However, there is limited synthesized knowledge on Software Engineering (SE) approaches for building, operating, and maintaining AI-based systems.
To collect and analyze state-of-the-art knowledge about SE for AI-based systems, we conducted a systematic mapping study.
We considered 248 studies published between January 2010 and March 2020.
SE for AI-based systems is an emerging research area, where more than 2/3 of the studies have been published since 2018. The most studied properties of AI-based systems are dependability and safety. We identified multiple SE approaches for AI-based systems, which we classified according to the SWEBOK areas. Studies related to software testing and software quality are very prevalent, while areas like software maintenance seem neglected. Data-related issues are the most recurrent challenges.
Our results are valuable for: researchers, to quickly understand the state-of-the-art and learn which topics need more research; practitioners, to learn about the approaches and challenges that SE entails for AI-based systems; and, educators, to bridge the gap among SE and AI in their curricula.
\end{abstract}




\maketitle

\section{Introduction}\label{sec:intro}
In the last decade, increased computer processing power, larger datasets, and better algorithms have enabled advances in Artificial Intelligence (AI) \cite{anthes2017artificial}. Indeed, AI has evolved towards a new wave, which Deng calls “the rising wave of Deep Learning” (DL) \cite{deng2018artificial}\footnote{\url{https://en.wikipedia.org/wiki/History_of_artificial_intelligence##Deep_learning,_big_data_and_artificial_general_intelligence:_2011-present}}. DL has become feasible, leading to Machine Learning (ML) becoming integral to many widely used software services and applications \cite{deng2018artificial}. For instance, AI has brought a number of important applications, such as image- and speech-recognition and autonomous, vehicle navigation, to near-human levels of performance \cite{anthes2017artificial}.

The new wave of AI has hit the software industry with the proliferation of AI-based systems integrating AI capabilities based on advances in ML and DL \cite{Amershi2019, Bosch2020}. AI-based systems are software systems which include AI components. These systems learn by analyzing their environment and taking actions, aiming at having an intelligent behaviour. As defined by the expert group on AI of the European Commission, “AI-based systems can be purely software-based, acting in the virtual world (e.g. voice assistants, image analysis software, search engines, speech and face recognition systems) or AI can be embedded in hardware devices (e.g. advanced robots, autonomous cars, drones or Internet of Things applications)”\footnote{\url{https://ec.europa.eu/digital-single-market/en/news/definition-artificial-intelligence-main-capabilities-and-scientific-disciplines}}.

Building, operating, and maintaining AI-based systems is different from developing and maintaining traditional software systems. In AI-based systems, rules and system behaviour are inferred from training data, rather than written down as program code \cite{khomh2018software}. AI-based systems require interdisciplinary collaborative teams of data scientists and software engineers \cite{Amershi2019}. The quality attributes for which we need to design and analyze are different \cite{ozkaya2020really}. The evolution of AI-based systems requires focusing on large and changing datasets, robust and evolutionary infrastructure, ethics and equity requirements engineering \cite{Kaestner2020}. Without acknowledging these differences, we may end up creating poor AI-based systems with technical debt \cite{Sculley2015}.

In this context, there is a need to explore Software Engineering (SE) practices to develop, maintain and evolve AI-based systems. This paper aims to characterize SE practices for AI-based systems in the new wave of AI, i.e., \textbf{Software Engineering for Artificial Intelligence (SE4AI)}. The motivation of this work is to synthesize the current SE knowledge pertinent to AI-based systems for: researchers to quickly understand the state of the art and learn which topics need more research; practitioners to learn about the approaches and challenges that SE entails when applied to AI-based systems; and educators to bridge the gap among SE and AI in their curricula.

Bearing this goal in mind, we have conducted a Systematic Mapping Study (SMS) considering literature from January 2010 to March 2020. The reason to focus on the last decade is that this new wave of AI started in 2010, with industrial applications of DL for large-scale speech  recognition, computer vision and machine translation \cite{deng2018artificial}.

The main contributions of this work are the synthesis of:

\begin{itemize}
\item Bibliometrics of the state of the art in SE4AI (see Section \ref{sec:results-rq1}).
\item Characteristics of AI-based systems, namely scope, application domain, AI technology, and key quality attribute goals (see Section \ref{sec:results-rq2})).
\item SE approaches for AI-based systems following the Knowledge Areas of SWEBOK, a guide to the
SE Body of Knowledge \cite{bourque2014swebok} (see Section \ref{sec:results-rq3})).
\item Challenges of SE approaches for AI-based systems following SWEBOK Knowledge Areas (see Section \ref{sec:results-rq4})).
\end{itemize}

\section{Background}\label{sec:background}

This section respectively discusses the synergies between SE and AI, and related work.

\subsection{On the synergies between SE and AI}

\paragraph{History of AI and SE}
AI and SE are naturally related, as they both have their roots in computer science. Although AI as we know it today may be traced back to the early 50s or 40s, software development and AI met for the first time when the Mark 1 perceptron was programmed on an IBM 704 machine for image recognition, in 1959 – although the perceptron itself was initially intended to be a machine and software development has rather little to do with SE as we know it today. SE as a profession and research area was developed in the late 1960s \cite{naur1969software}, when the term “software crisis” was coined relating to problems in engineering increasingly large and complex software systems. The boom of Expert Systems in the 80s brought AI back to the forefront again \cite{Russell.2021}. The deployment and use of expert systems in production systems has naturally led to a series of SE challenges (like validation \& verification).  For example, Partridge \cite{partridge1987methodology} surveyed the relationships between AI and SE. Despite addressing problems of different nature, AI and SE have been intertwined since their very beginnings: AI methods have been used to support SE tasks (AI4SE) and SE methods have been used to develop AI (SE4AI) software.

\paragraph{AI4SE}
Recently, Perkusich et al. \cite{perkusich2020intelligent} referred to AI4SE as intelligent SE and defined it as a portfolio of SE techniques, which “explore data (from digital artifacts or domain experts) for knowledge discovery, reasoning, learning, planning, natural language processing, perception or supporting decision-making”. AI4SE has developed driven by the rapid increase in size and complexity of software systems and, in consequence, of SE tasks. Wherever software engineers came to their cognitive limits, automatable methods were the subject of research. While searching for solutions, the SE community observed that a number of SE tasks can be formulated as data analysis (learning) tasks and thus can be supported, for example, with ML algorithms. 

\paragraph{SE4AI}
First applications of SE to AI were limited to simply implementing AI algorithms as standalone programs, such as the aforementioned Mark 1 perceptron. As AI-based software systems grew in size and complexity and as its practical and commercial application increased, more advanced SE methods were required. The breakthrough took place when AI components became a part of established software systems, such as expert systems, or driving control. It quickly became clear that, because of the specific nature of AI (e.g., dependency on learning data), traditional SE methods were not suitable anymore (e.g., leading to technical debt \cite{Sculley2015}). This called for revision of classical, and development of new, SE paradigms and methods. This paper provides a comprehensive review of what has been achieved in the area so far.

\subsection{Related work on SE4AI}
Several secondary studies in the broad area of SE4AI have been published so far (see Table \ref{tab:related-work}).

Masuda et al. \cite{masuda2018survey} conducted a review to identify techniques for the evaluation and improvement of the software quality of ML applications. They analyzed 101 papers and concluded that the field is still in an early state, especially for quality attributes other than functional correctness and safety.

Washizaki et al. \cite{washizaki2019studying} conducted a multivocal review to identify architecture and design patterns for ML systems. From 35 resources (both white and grey literature), they extracted 33 unique patterns and mapped them to the different ML phases. They discovered that, for many phases, only very few or even no patterns have been conceptualized so far. Serban and Visser \cite{serban2021empirical} performed both a case study and a systematic literature review on the topic of software architecture for machine learning. They reviewed 42 studies and performed 10 semi-structured interviews with practitioners from 10 different organisations. In their paper, the authors report 20 challenges and potential solutions.
On a similar topic, John et al. \cite{john2020architecting}, performed a systematic review of both scientific (13 studies) and grey literature (6 studies) on the topic of deployment of ML systems. They report a total of 27 challenges and 52 practices.
Lorenzoni et al. \cite{lorenzoni2021machine} performed a systematic literature review on the topic of development of machine learning systems. They analysed 33 studies between 2010 and 2020 and classified 10 issues and 13 solutions into seven SE practices.

A number of reviews have been conducted in the area of software testing. Borg et al. \cite{borg2018safely} performed a review of verification and validation techniques for deep neural networks (DNNs) in the automotive industry. From 64 papers, they extracted challenges and verified them with workshops and finally a questionnaire survey with 49 practitioners. They conclude, among other challenges, that a considerable gap exists between safety standards and nature of contemporary ML-based systems.
Another study was published by Ben Braiek and Khomh \cite{braiek2020testing}. In their review of testing practices for ML programs, they selected a total of 37 primary studies and extracted challenges, solutions, and gaps. The primary studies were assigned to the categories of detect errors in data (five papers), in ML models (19 papers), and in the training program (13 papers).
Riccio et al. \cite{riccio2020testing} extracted testing challenges from 70 primary studies and propose the following categories: realism of test input data (5 papers), adequacy of test criteria (12 papers), identification of behavioural boundaries (2 papers), scenario specification and design (3 papers), oracle (13 papers), faults and debugging (8 papers), regression testing (5 papers), online monitoring and validation (8 papers), cost of testing (10 papers), integration of ML models (2 papers), and data quality assessment (2 papers). 
Similarly, Zhang et al. \cite{zhang2020machine} surveyed the literature on ML testing and selected 138 papers. From these, they summarized the tested quality attributes (e.g. correctness or fairness), the tested components (e.g. the data or the learning program), workflow aspects (e.g. test generation or evaluation), and application contexts (e.g. autonomous driving or machine translation).

In addition to these specialized reviews, there are also some secondary studies more similar to ours, i.e. that have a general SE focus. Serban et al. \cite{serban2020adoption} conducted a multivocal review with 21 relevant documents (both white and grey literature) to identify and analyze SE best practices for ML applications. They extracted 29 best practices and used a follow-up questionnaire survey with 313 software professionals to find out the degree of adoption and impact of these practices.
Furthermore, Wang et al. \cite{wang2020synergy} took a broader view and conducted a systematic literature review about general synergies between ML/DL and SE, i.e. covering both machine learning for SE (ML4SE) and SE for machine learning (SE4ML) research. However, only 15 of the 906 identified studies covered the SE4ML direction. Based on their results, the authors concluded that “it remains difficult to apply SE practices to develop ML/DL systems”.
Another systematic literature review was performed by Kumeno \cite{kumeno2019sofware}. He focused solely on the extraction of SE challenges for ML applications and mapped them to the different SWEBOK areas. In total, he selected 115 papers (47 papers focusing on SE-related challenges for ML and 68 papers focusing on ML-techniques or ML-applications challenges) from 2000 to 2019.
Moreover, Lwakatare et al. \cite{lwakatare2020large} conducted a similar review of SE challenges faced by industrial practitioners in the context of large-scale ML systems. They categorized 23 challenges found in the 72 papers selected according to four quality attributes (adaptability, scalability, privacy, safety) and four ML process stages (data acquisition, training, evaluation, deployment). Adaptability and scalability were reported to face a significantly larger number of challenges than privacy and safety. They also identified 8 solutions, e.g. transfer learning and synthetically generated data, solving up to 13 of the challenges, especially adaptability. 
Giray \cite{giray2020software} also conducted a systematic literature review to identify the state of the art and challenges in the area of ML systems engineering. In his sampling method, he exclusively targeted publications from SE venues and selected 141 studies. These studies were then analyzed for their bibliometrics, the used research methods, plus mentioned challenges and proposed solutions.
Lastly, Nascimento et al. \cite{nascimento2020software} performed an SLR to analyze how SE practices have been applied to develop AI or ML systems, with special emphasis on limitations and open challenges. They also focus on system contexts, challenges, and SE contribution types. While they considered publications between 1990 and 2019, they only selected 55 papers.

We summarize in Table \ref{tab:related-work} the findings of the related work. This summary has a three-fold objective: (i) to provide a synthetic view of the approaches aforementioned; (ii) to make evident our claim that no systematic mapping or general review with a breadth and depth similar to ours has been published so far; (iii) to facilitate the comparison of the results of our study with the related work.

In summary, even though several secondary studies have recently been published or submitted (a few of the aforementioned studies are pre-prints), there are still very few works that broadly summarize and classify research in the field of SE for AI-based systems. No systematic mapping or general review with a breadth and depth similar to ours has been published so far. Existing general reviews focus either exclusively on challenges, analyze considerably fewer studies, or only take publications with an industry context into account, i.e. they partially fail to describe the wide spectrum of results in this research area.

\begin{small}
\begin{longtable}{|p{0.65cm}|p{1.65cm}|p{3.25cm}|p{4.5cm}|p{3.5cm}|}
    %\centering
    \caption{Summary of relevant aspects on bibliometrics, AI-based system properties, SE approaches, and challenges found in related work.} \label{tab:related-work} \\
      %\small %\begin{tabular}{|p{0.65cm}|p{1.65cm}|p{3.25cm}|p{4.5cm}|p{3.5cm}|}
            \hline
            \textbf{Study} & \textbf{Bibliometrics} & \textbf{AI-based systems properties} & \textbf{SE approaches} & \textbf{Challenges} \\
            \hline
            \cite{masuda2018survey} & 101 studies (2005-2018) & safety, correctness &   practices for the evaluation and improvement of the software quality of ML applications & ML quality assurance\\ % Justus ML
            \hline
            \cite{washizaki2019studying} & 38 studies (2008-2019) & & 33 unique architecture and design patterns for ML systems & \\ % Silverio ML
            \hline
            \cite{borg2018safely} & 64 studies (2002-2007, 2013-2016) & safety, robustness, reliability & verification and validation techniques for safety-critical automotive systems & verification and validation in DNNs\\ % Justus DL
            \hline
            \cite{braiek2020testing} & 37 studies (2012-2018) & correctness & testing practices & 18 challenges organized in 6 dimensions: implementation issues, data issues, model issues, written code issues, execution environment issues, mathematical design issues\\  % Anna Maria ML
            \hline
            \cite{riccio2020testing} & 70 studies (2004-2019) & fairness, accuracy, safety, consistency & functional testing, test case generation and test oracle, integration testing, system testing & 11 challenges\\  % Adam ML
            \hline
            \cite{zhang2020machine} & 138 studies (2007-2019) & correctness, model relevance, robustness, security, data privacy, efficiency, fairness, interpretability &  testing workflow, testing components, testing properties, application scenarios & 4 testing challenges categories: test input generation, test assessment criteria, oracle problem, testing cost reduction \\ % Julien ML
            \hline
            \cite{serban2020adoption} & 21 studies (2017-2019) & & 29 best practices for ML systems in six categories: data, training, coding, deployment, team, governance & \\ % Silverio ML
            \hline
            \cite{wang2020synergy} & 15 studies (out of 906) (2009-2018) & & Model Evaluation, Deployment &\\
            \hline
            \cite{kumeno2019sofware} & 115 studies (2003-2019) & safety, security, ethics and regulation, software structure, testability, maintainability, performance, risk and uncertainty, economic impact & all SWEBOK areas & SE challenges for ML applications\\ % Marc ML
            \hline
            \cite{lwakatare2020large} & 72 studies (1998-2018) & adaptability, scalability, safety, privacy & Software construction. Software maintenance & 23 SE challenges faced by practitioners in the context of large-scale ML systems. 13 of them had solutions\\ % Xavi ML
            \hline
            \cite{giray2020software} & 141 studies (2007–2019) & & requirements engineering, design, software development and tools, testing and quality, maintenance and configuration management, SE process and management, organizational aspects & 31 challenges and partially solutions that have been raised by SE researchers\\ % Stefan ML
            \hline
            \cite{nascimento2020software} & 55 studies (1999-2019) & AI ethics, interpretability, scalability, explanation on failure, complexity, efficiency, fairness, imperfection, privacy, safety, safety and stability, robustness, reusability, stability, staleness & 
            AI software quality, data management, project management, infrastructure, testing, model development, requirement engineering, AI engineering, architecture design, model deployment, integration, education, operation support. & SE challenges for ML applications and how SE practices adjusted to deal with them.\\ % Marc ML
            \hline
            \cite{lorenzoni2021machine} & 33 studies (2010-2020) & & data processing, documentation and versioning, non-functional requirements, design and implementation, evaluation, deployment and maintenance, software capability maturity model & 10 issues and 13 solutions\\ % Julien
            \hline \cite{serban2021empirical} & 42 studies (2016-2021) & Performance, scalability, interpretability, hardware resources, interoperability, robustness, generalization, low data quality, scarcity of data, maintainability, privacy, security & Requirements, data, design, testing, operations, organisation & 20 challenges and potential solutions \\
            \hline \cite{john2020architecting} & 19 studies (2017-2020) & & Design, integration, deployment, operation, evolution. & 27 challenges and 52 practices \\
            \hline
            This study & 248 studies (2010-2020) & 40 quality attributes (see sections \ref{sec:results-rq1} and \ref{sec:results-rq2}) & 11 SWEBOK areas (see Section \ref{sec:results-rq3}) & 94 challenges (see Section \ref{sec:results-rq4})\\
            \hline
 %           This study & 248 studies (2010-2020) Discussed 6 observations &
 %           Classified 248 studies in three AI-based systems dimensions: scope, domain, and AI technologies
 %           
 %           Classified 190 studies regarding addressed quality attributes at high, medium, and low level extending ISO 25000
 %           
 %           Discussed 5 observations & ML, DL, other &
 %           
 %           Revealed SE approaches for AI-based systems from 248 studies mapped into SWEBOK areas, identifying 8 important ones
 %           
 %           Discussed 4 observations &
 %           Revealed challenges in AI-based systems in SWEBOK areas from 39 studies, proposing new categories inside knowledge areas
 %           
 %           Discussed 3 observations \\
 %           \hline
        %\end{tabular}
        
\end{longtable}
\end{small}

The main objective of our study is therefore to systematically select and review the broad body of literature and to present a holistic overview of existing studies in SE for AI-based systems. An SMS is a proven and established method to construct such a holistic overview.

\section{Research Methodology}\label{sec:methodology}
This SMS has been developed following the guidelines for performing SMSs from Petersen et al. \cite{petersen2015guidelines}. Additionally, the guidelines for systematic literature reviews provided by Kitchenham and Charters \cite{kitchenham2007guidelines} have also been used when complementary. This is because the method for searching for the primary studies, and taking a decision about their inclusion or exclusion is very similar between a systematic literature review and an SMS. This SMS has five main steps \cite{petersen2008systematic, kuhrmann2017} described in the following subsections.

\subsection{Definition of research questions}
The aim of this SMS consists of identifying the existing research of SE for AI-based systems by systematically selecting and reviewing published literature, structuring this field of interest in a broader, quantified manner. We define the goal of this SMS formally using GQM \cite{basili1994gqm}: \textit{Analyze} SE approaches \textit{for the purpose of} structure and classification \textit{with respect to} proposed solutions and existing challenges \textit{from the viewpoint of} both SE researchers and practitioners \textit{in the context of} AI-based systems in the new wave of AI.

To further refine this objective, we formulated four Research Questions (RQs) to be answered by our primary studies:
\begin{itemize}
    \item[RQ1.] How is SE research for AI-based systems characterized? \label{RQ1}
    \item[RQ2.] What are the characteristics of AI-based systems (used terms, scope, and quality goals)? \label{RQ2}
    \item[RQ3.] Which SE approaches for AI-based systems have been reported in the scientific literature? \label{RQ3}
    \item[RQ4.] What are the existing challenges associated with SE for AI-based systems? \label{RQ4}
\end{itemize}

The research methodology followed to answer each of these RQs is detailed in Section \ref{sec:methodology-analysis}.

\subsection{Conduct search}
Since AI-based systems cover interdisciplinary communities (SE and AI), it is difficult to find an effective search string. Therefore, we decided to execute a hybrid search strategy \cite{mourao2020}, applying both a search string in Scopus\footnote{\url{https://www.scopus.com}} and a snowballing strategy \cite{wohlin2014}. This strategy is referred to as \textbf{“Scopus + BS*FS”} \cite{mourao2020}: it first runs a search over Scopus to get a start set of papers and compose a seed set for snowballing. The Scopus database contains peer-reviewed publications from top SE journals and conferences, including IEEE Xplore, ACM Digital library, ScienceDirect (Elsevier), and Springer research papers. We have recently assessed that Scopus’ coverage is optimal when compared to these other databases \cite{Costal2021}. Scopus enabled us to search with one engine for relevant studies considering a large library content and using different search engine functionalities such as export of the results and formulation of own search strings. We iteratively created several initial search strings, discussed their containing terms, and compared the results. The final search string used on Scopus was: 

\begin{itemize}
    \item[] TITLE-ABS-KEY("software engineering for artificial intelligence" OR
    \item[] "software engineering for machine learning" OR
    \item[] (("software engineering" OR "software requirement*" OR "requirements engineering" OR "software design" OR "software architecture" OR "software construction" OR "software testing" OR "software maintenance" OR "software configuration management" OR "software quality")
    \item[] \begin{center} AND \end{center}
    \item[] ("AI-enabled system*" OR "AI-based system*" OR "AI-infused system*" OR "AI software" OR "artificial intelligence-enabled system*" OR "artificial intelligence-based system*" OR "artificial intelligence-infused system*" OR "artificial intelligence software" OR "ML-enabled system*" OR "ML-based system*" OR "ML-infused system*" OR "ML software" OR "machine learning-enabled system*" OR "machine learning-based system*" OR "machine learning-infused system*" OR "machine learning software"))) 
\end{itemize}

With this search string, we aimed to capture the literature on SE4AI. Since this is a concept recently popular in the SE community, it was directly used as part of the search string. Furthermore, we also included subareas of AI, such as ML. Finally, the search string also considered the combination of SE Knowledge Areas from SWEBOK v3.0 \cite{bourque2014swebok} and many similar terms to AI-based systems. We ran a check whether this search string included key papers on the topic of SE4AI from an annotated bibliography of an (external) expert on the topic \footnote{\url{https://github.com/ckaestne/seaibib}}, and many were included. Therefore, we considered it as a good starting point to be complemented with other techniques to mitigate selection bias.\\

\begin{figure}
  \centering
    \includegraphics[width=\textwidth]{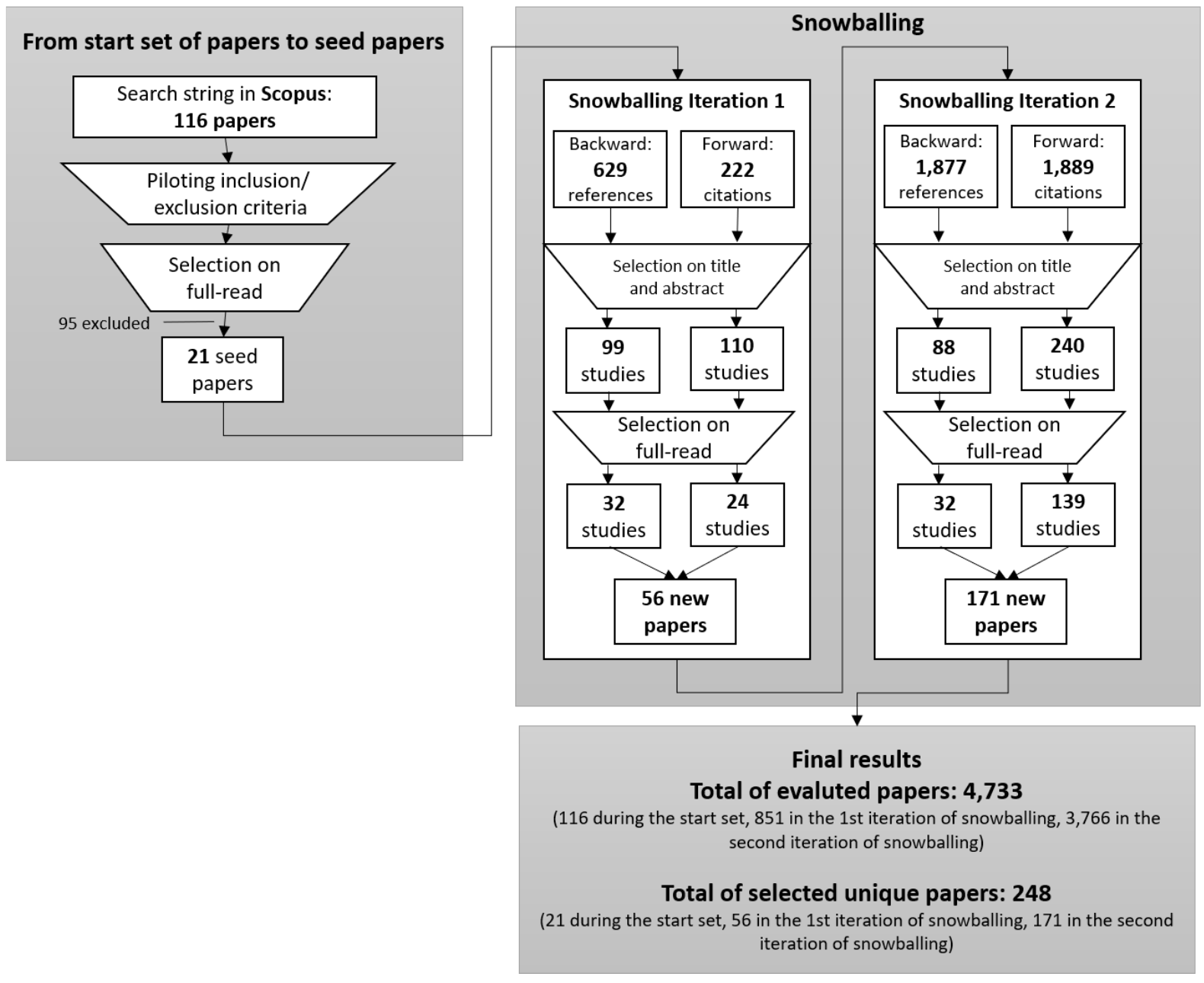}
    \caption{Search process (selection on title and abstract appear in the snowballing file, new papers are in the data extraction).}
    \label{fig:Figure_3_1}
\end{figure}

An important decision was to only consider primary studies belonging to the new wave of AI from January 2010 to March 2020. The reason is that we wanted to structure the knowledge of SE4AI since the new wave of AI \cite{anthes2017artificial, deng2018artificial}. We applied the search string on Scopus to this interval of time on April 2\textsuperscript{nd}, 2020, which resulted in 116 studies (see Figure \ref{fig:Figure_3_1}).

Then, other papers were obtained from the seed set via backward and forward snowballing. We applied snowballing as indicated in the guidelines by Wohlin \cite{wohlin2014}. The index used to check the number of references was Google Scholar. While we considered all the citations of each paper during backward snowballing, we established a limit of the first 100 citations returned by Google Scholar during forward snowballing. Each of the authors  checked these references and citations of randomly assigned papers. To avoid missing relevant papers, if there was any hint that a study could be included, it was inserted in our snowballing working document.

In total, we performed two snowballing iterations as reported in this study. For each entry in the aforementioned snowballing working document, at least two researchers applied inclusion and exclusion criteria. The next subsection explains this selection process of screening papers.

\subsection{Screening of papers}
\begin{table}
\caption{Inclusion and exclusion criteria.}
\begin{tabularx}{\textwidth}{|l|X|}
\hline
\textbf{Criteria} & \textbf{Description} \\ \hline
Inclusion (IC1) & The work is mainly focused on SE for AI-based systems. \\ \hline
Exclusion (EC1) & The work does not fulfill IC1 (e.g., is focused on AI technologies for SE). \\ \hline
Exclusion (EC2) & The work is not accessible even after contacting the authors. \\ \hline
Exclusion (EC3) & The work is not completely written in English. \\ \hline
Exclusion (EC4) & The work is a secondary study (e.g., SLR). \\ \hline
Exclusion (EC5) & The work is an exact duplicate of another study. \\ \hline
Exclusion (EC6) & The work is a short paper of two pages or less. \\ \hline
Exclusion (EC7) & The work is not a research paper published in books, journals, conferences, workshops, or the arXiv repository (e.g., an editorial for a special issue, a table of contents of proceedings, short course description, tutorial, summary of a conference, Ph.D. thesis, master thesis, blog, technical report).\\ \hline
Exclusion (EC8) & The work has been published before January 2010 or after March 2020. \\ \hline
\end{tabularx}
\label{tab:Tab_3_1}
\end{table}

Based on our RQs, we specified the inclusion and exclusion criteria listed in Table \ref{tab:Tab_3_1}. This supported a consistent screening among all of us. %Furthermore, we screened each paper by at least two persons, permuted the assignments of pairs of persons, and discussed  the applied criteria to clarify any open question.
These criteria had been used to select relevant papers first based on their titles and abstracts and afterward based on the full-reads. We received English titles and abstracts for all the obtained studies, so there are no exclusions based on the language of the primary study. Knowledge of the authors, institutions, journals, and year of publication had not been removed during the study selection process. Evidence suggests that masking the origin of primary studies does not improve reviews \cite{kitchenham2004procedures}. As we started the search for primary studies beginning of April 2020, we excluded papers that were published after March 2020 but identified through forward snowballing.\\

For screening each paper identified both in the search in Scopus and the snowballing, we assigned two researchers who checked the abstract and title independently of each other. We made sure to mix the assignments of persons so that each person had a similar amount of primary studies with everybody from the research team, and permuting the assignments, so that pairs were balanced. In case of disagreement, a third person assisted to find a common decision. This aimed to improve the reliability of our systematic mapping \cite{wohlin2013}. The inter-rater agreement before the third person was involved was 0.751 using Cohen's kappa coefficient, which indicates a substantial agreement among participants~\cite{Emam1999}. The full-read was done by one person and if the final decision mismatched the previous agreement, all included persons were informed to eventually decide about the inclusion or exclusion of the corresponding paper. We documented these results in our snowballing working document to ease the transparency among us. Following this, we identified in total 21 relevant seed papers, 56 additional papers based on the first snowballing iteration, and additional 171 papers in our second snowballing iteration as illustrated in Figure \ref{fig:Figure_3_1}. In 53 cases (9.83\%), a third person was needed to resolve the disagreements. After finishing the second snowballing iteration, we included 248 primary studies in total. As the field of SE4AI is further emerging and our resources for this study are limited, we decided to stop at this point.

\subsection{Keywording}
We created a data extraction form containing four different kinds of information to gain a broad view of SE for AI-based systems related to our RQs. We planned to collect the following data:
\begin{enumerate}
    \item Generic data and bibliometrics: contains demographic information as well as the used criteria to assess the quality of our selected studies.
    \item AI-related terms: includes terms for the targeted AI-based system or subsystem/component of the AI-based system as used in the primary studies, addressed AI-based system properties (e.g., explainability, safety, etc.) as focused by the primary studies, definitions of the terms of the targeted AI-based system, and domain names as reported in the primary studies.
    \item SE approaches for AI-based systems: contains the most suitable Knowledge Areas of the reported SE approaches for AI-based systems, the SE approach name or description categorized according to different types of approaches (method, model, practice, tool, framework, guideline, other), and the observed impact of applying the SE approach.
    \item Challenges of SE approaches for AI-based systems: list of explicitly stated challenges reported in the primary studies and, if available, possible solutions and relevant quotes regarding SE approaches for AI-based systems.
\end{enumerate}

For our data extraction form, we made use of existing classification criteria coming from the SE domain. More precisely, we applied Ivarsson and Gorscheck's rigor and relevance quality assessment model \cite{ivarsson2011method} for the quality appraisal (see Section \ref{sec:characts}), and we used the Knowledge Areas listed in the SWEBOK v3.0 \cite{bourque2014swebok} to classify the identified SE approaches for AI-based systems. These are the corresponding Knowledge Areas: Software Requirements, Software Design, Software Construction, Software Testing, Software Maintenance, Software Configuration Management, Software Engineering Management, Software Engineering Process, Software Engineering Models and Methods, Software Quality, Software Engineering Professional Practice, Software Engineering Economics, Computing Foundations, Mathematical Foundations, Engineering Foundations.

The values of some criteria were fixed in advance (e.g., the aforementioned quality appraisal criteria values and the predefined Knowledge Areas), whilst others emerged in the analysis phase after reading the papers (e.g., AI-related terms, focused quality properties).

We piloted the initial data extraction form to both ensure a common understanding among all the involved researchers and extend the form if required, especially with additional values for the classification criteria. For this piloting activity, we randomly selected three of our seed papers, fully read the papers, extracted them independently by the eight authors, and discussed our results together. As a consequence, we improved the extraction form to make a few columns more objective and decided to use literal quotations for some columns (e.g. challenges). Because we expected that the topic of our study is quite diverse, we also decided to perform the data extraction based on full-read. Just checking the abstracts, introductions, and conclusions would not contain enough information to create the final keywording, i.e., the classification criteria with several values. Furthermore, other researchers reported that a quick keywording relies on good quality of the abstracts, and in the SE domain they may not be enough to conduct the keywording \cite{petersen2008systematic, brereton2007}.

During the discussions of our data extraction form, we observed that the used quality appraisal classification criteria by Ivarsson and Gorscheck are missing specific values for better capturing the AI-based systems or the study types used in the AI/ML community. For example, we identified several primary studies that compared the performance and outcome of an AI/ML system with other systems or under different environments. As we count this as a type of empirical study, we added a new method type named \textit{Benchmark}. Primary studies belong to this kind of study type if they describe a rigorous study that evaluates one or more algorithms in well-established operational settings (with data and variables). Furthermore, we extended the definitions of the criteria \textit{Industrial Relevance} by including AI/ML algorithms, systems, and data sets as another type of subject. This allowed us to assess the industrial relevance of Benchmark studies.

\subsection{Data extraction and mapping process}\label{sec:methodology-analysis}
Each of the 248 primary studies was assigned to a single researcher for extraction based on the predefined data extraction form. Extractors were in frequent asynchronous contact to discuss potential inconsistencies. The weekly project meeting was also used for synchronization on this matter. In these meetings, we discussed the most persistent conflicts and shared our way of working as well as emerging findings to ensure cohesion of the team and to minimize subjectivity. Additionally, the data collection form includes the name of the reviewer and space for additional notes. This enabled us also to keep track of this process. 

After completing the extraction, we started data analysis and synthesis. In general, we performed both quantitative and qualitative analyses to classify the extracted data. During this process, additional extraction inconsistencies and mistakes were discovered and easily resolved in direct communication with the original extractors. Synthesis per RQ was performed by groups of at least two researchers, who often re-read (parts of) the original papers for this and kept the group in the loop. Final results were presented to the rest of the team and feedback was incorporated. 

Regarding the required mapping and analysis to answer the RQs, we performed the following activities in each RQ: 

\textbf{For the analysis of RQ1}, we used the assessed rigor and relevance quality \cite{ivarsson2011method} and performed frequency analysis to additionally determine bibliographical data such as the annual publication trend, venue types, authors’ affiliations, geography distribution, and the empirical research type of the primary studies.

\textbf{To answer RQ2}, we counted the number of occurrences of various terms related to AI used to characterize the study objects. We then used inductive coding to distill dimensions (Scope, Application Domain, and Technologies of AI) to describe the study objects, coded all primary studies, and reported frequencies. For the key quality attribute goals of AI-based systems, this also included a harmonization of the used terms as well as axial coding to cluster the identified quality properties.

\textbf{As RQ3 and RQ4 highly rely on extensive qualitative analysis}, all eight authors supported the analysis. Therefore, we split the primary studies according to the extracted SWEBOK Knowledge Areas and equally distributed the number of studies to groups of two or three researchers. Again, we performed thematic analysis \cite{cruzes2011recommended} and derived several codes independently for the different Knowledge Areas (RQ3) and uniformed them to our subcategories for each of the SWEBOK Knowledge Areas. Finally, we mapped the primary studies accordingly, whereby one primary study could belong to several Knowledge Areas. For each Knowledge Area, each primary study could also be mapped to several subcategories, but required at least one.

As we created a subcategory called “challenges” for each Knowledge Area, we restricted the analysis for RQ4 regarding challenges in SE4AI explicitly to those primary studies whose classification under RQ3 included a mapping to this subcategory. For the mapping of the challenges, we made use of the different Topics provided for each Knowledge Area by SWEBOK. We mapped a challenge into one or more SWEBOK Topics if the link was evident. Moreover, we added information about root causes, mitigation actions, and impact on the data extraction form as a basis for some additional qualitative analysis. Finally, we uniformed the terminology (near 80\% of the Topics were rewritten to some extent) and the classification (changes of SWEBOK Topic or even Knowledge Area, assignment of an additional Topic to a challenge, or removal of a Topic). It also resulted in removing some challenges because they are duplicated in another primary study with proper citation (e.g., \cite{McDermid2019} presenting the same four challenges as \cite{Amodei2016}) or are too generic to allow proper analysis.
The three researchers responsible for RQ4 decided to apply inductive coding and added a new Knowledge Area called \textit{Software Runtime Behaviour} that they considered not well-covered in SWEBOK. Our new Knowledge Area was finally composed of three Topics: \textit{Cost \& Efficiency}, \textit{Quality}, and \textit{Robustness \& Operability}. Furthermore, seeking conceptual consistency inside some SWEBOK Topics, we proposed the following new Topics:
\begin{itemize}
    \item \textit{ML/AI methods}, in SE Models and Methods Knowledge Area. This new Topic could be numbered as 9.4.5, i.e., at the same level as existing particular types of methods, e.g., Formal Methods (9.4.2) or Agile Methods (9.4.4).
    \item \textit{ML/AI specific activities}, in Software Life Cycles Topic (8.4.2). The reason is that while these activities are part of ML/AI-based software development, the current formulation of the Topic does not explicitly address the definition of concrete activities in the life cycle.
    \item \textit{Data-related issues} and \textit{Process-related issues}, as sub-Topics in Software Testing: Key Issues (4.1.2). We considered it convenient to clearly distinguish among these two types of issues related to testing due to their different nature.
    \item \textit{Applicability and adoption of research results in practice}, as sub-Topic of SE Professional Practice. We thought that this important aspect is not covered by any of the three main Topics in the Knowledge Area (Professionalism, Group Dynamics and Psychology, and Communication Skills). Therefore, this proposed Topic could be numbered as 11.4.
\end{itemize}

In the replication package\footnote{Replication package available on \url{https://doi.org/10.6084/m9.figshare.14538324}.}, the reader can see: the results of the search on Scopus for seed papers (and alternative discarded search procedures), the result of the application of inclusion and exclusion criteria for the papers found, the document used for snowballing, the data extraction form, and analysis files for each RQ.

\section{RQ1: How is software engineering research for AI-based systems characterized?}\label{sec:results-rq1} 

This section respectively discusses the bibliometrics of the primary studies, and their research rigor and industrial relevance.

\subsection{Bibliometrics}

We focus the bibliometric analysis into four aspects: annual trend, distribution per venue type, distribution per affiliation type and geographical distribution.

\begin{figure}
  \centering
    \includegraphics[width=0.6\textwidth]{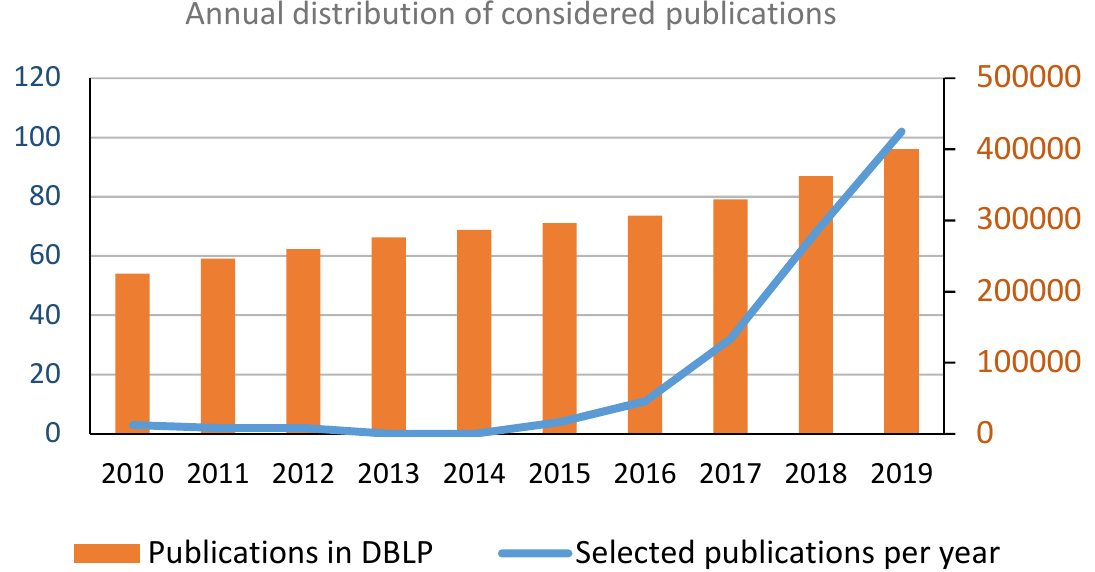}
    \caption{Annual distribution of publications.} \label{fig:Figure_4_1}
\end{figure}

\textbf{Annual trend.} We observe a first period in which the number of publications were marginal (2010-2012) or even nonexistent (2013-2014) (see Figure~\ref{fig:Figure_4_1}). But since 2015, we observe a rapidly increasing growth in the field, with the number of publications being approximately doubled every year since (from 4 in 2015 to 102 in 2019). This trend is well above the overall trend of increasing number of publications in DBLP, shown in the same figure for comparison.

The numbers for the year 2020 are not shown, as this year is not completely covered in the study (the timespan of the study covers until March 2020). 

\begin{figure}
  \centering
    \includegraphics[width=0.6\textwidth]{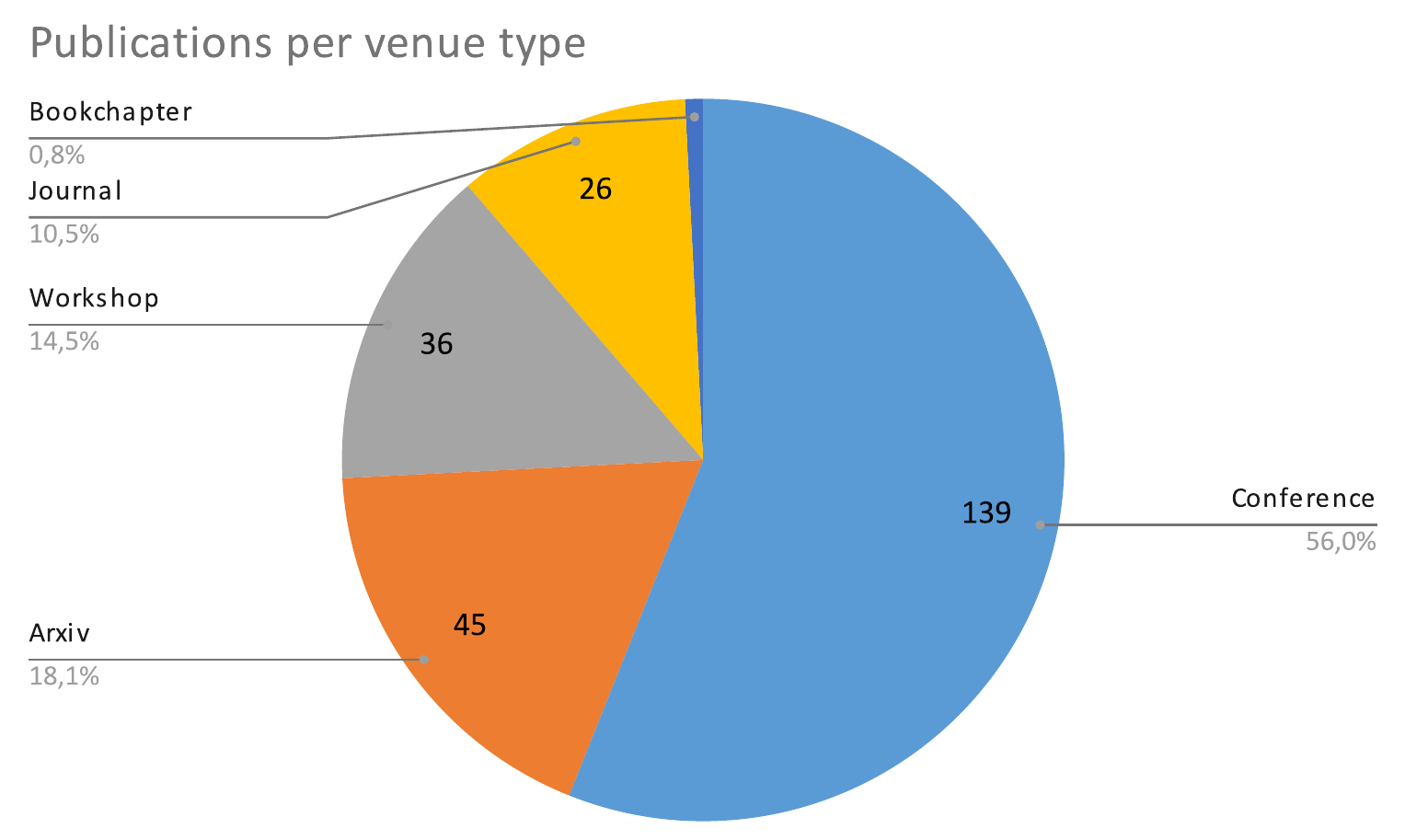}
    \caption{Number of publications per venue.} \label{fig:Figure_4_2}
\end{figure}

\textbf{Distribution per venue type.} As shown in Figure~\ref{fig:Figure_4_2}, most of the primary studies were published in conferences, and remarkably, there is also a considerable number of publications which were published as arXiv reports.

\begin{figure}
  \centering
    \includegraphics[width=0.6\textwidth]{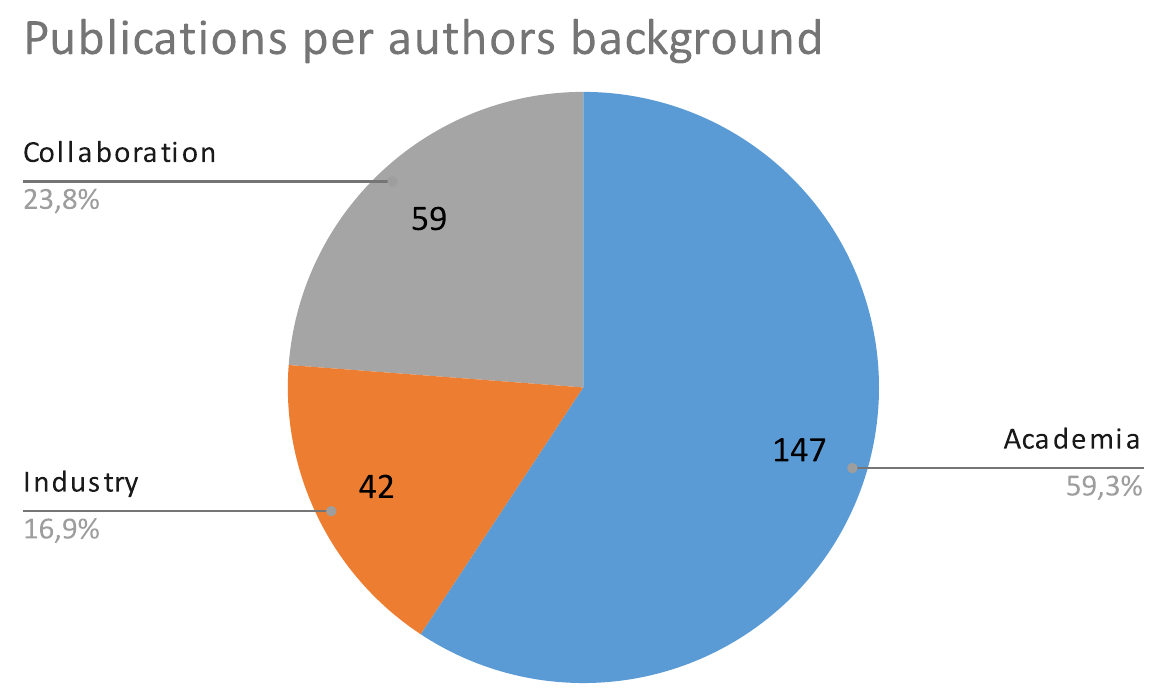}
    \caption{Number of publications per author affiliation.} \label{fig:Figure_4_3}
\end{figure}

\textbf{Distribution per affiliation type.} Figure~\ref{fig:Figure_4_3} depicts the type of affiliation of the authors of the research papers analysed. Academia includes authors with affiliation in research centers. 

\begin{figure}
  \centering
    \includegraphics[width=0.95\textwidth]{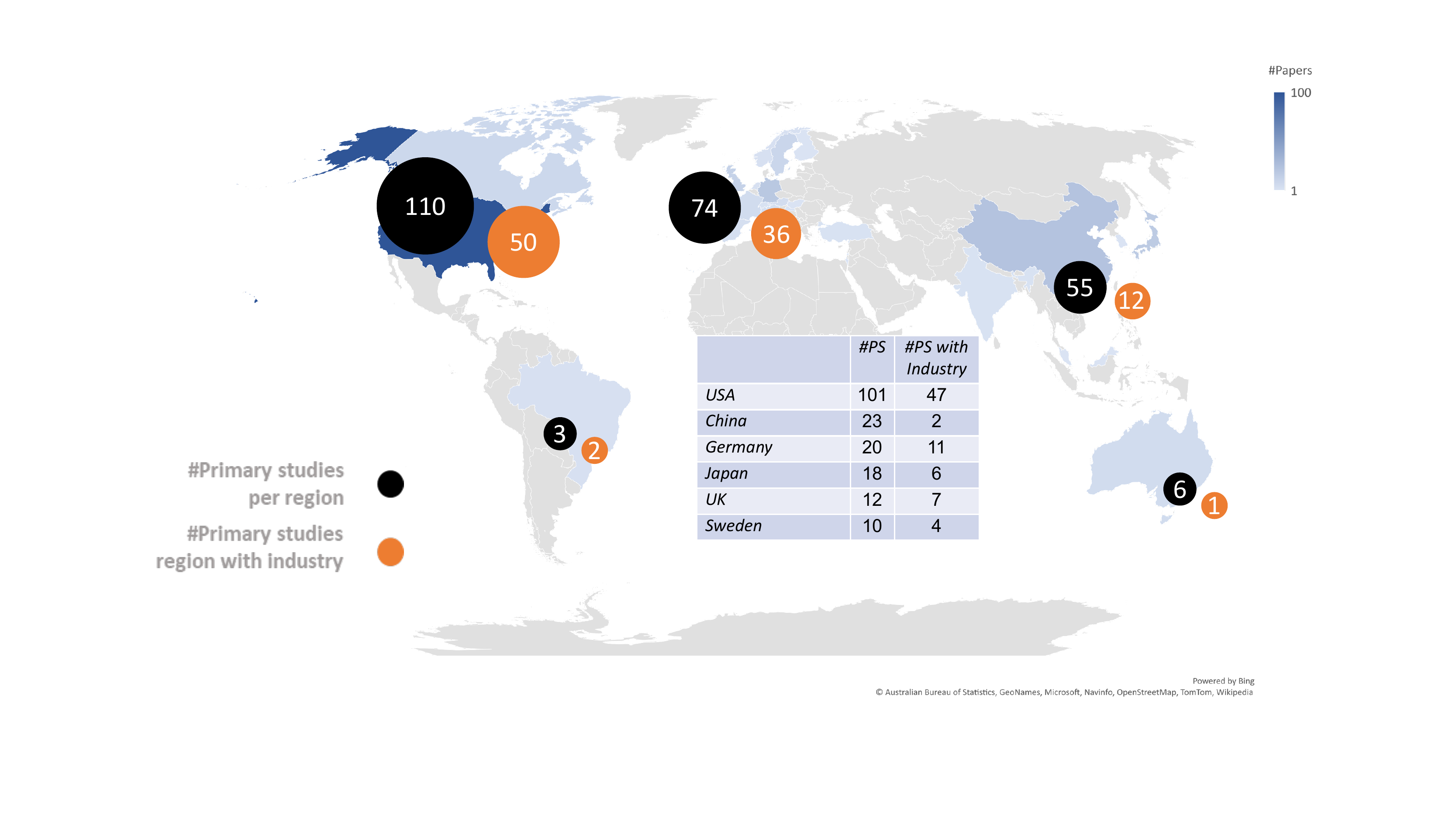}
    \caption{Distribution of publications by continent and country.} \label{fig:Figure_4_4}
\end{figure}

\textbf{Geographic distribution.} We examined the affiliation country of the first author of the papers (see Figure~\ref{fig:Figure_4_4}). As shown, the USA plays a leading role in 40\% of the studies analysed, which multiplies China's leading role by a 4x factor and Germany and Japan by 5x. If we observe the distribution by continents, North America doubles Asia in absolute numbers and the distance grows if we consider papers with authors in industry; Europe lies in the middle, with the rest of the continents with little or no presence.

\begin{figure}
  \centering
    \includegraphics[width=0.95\textwidth]{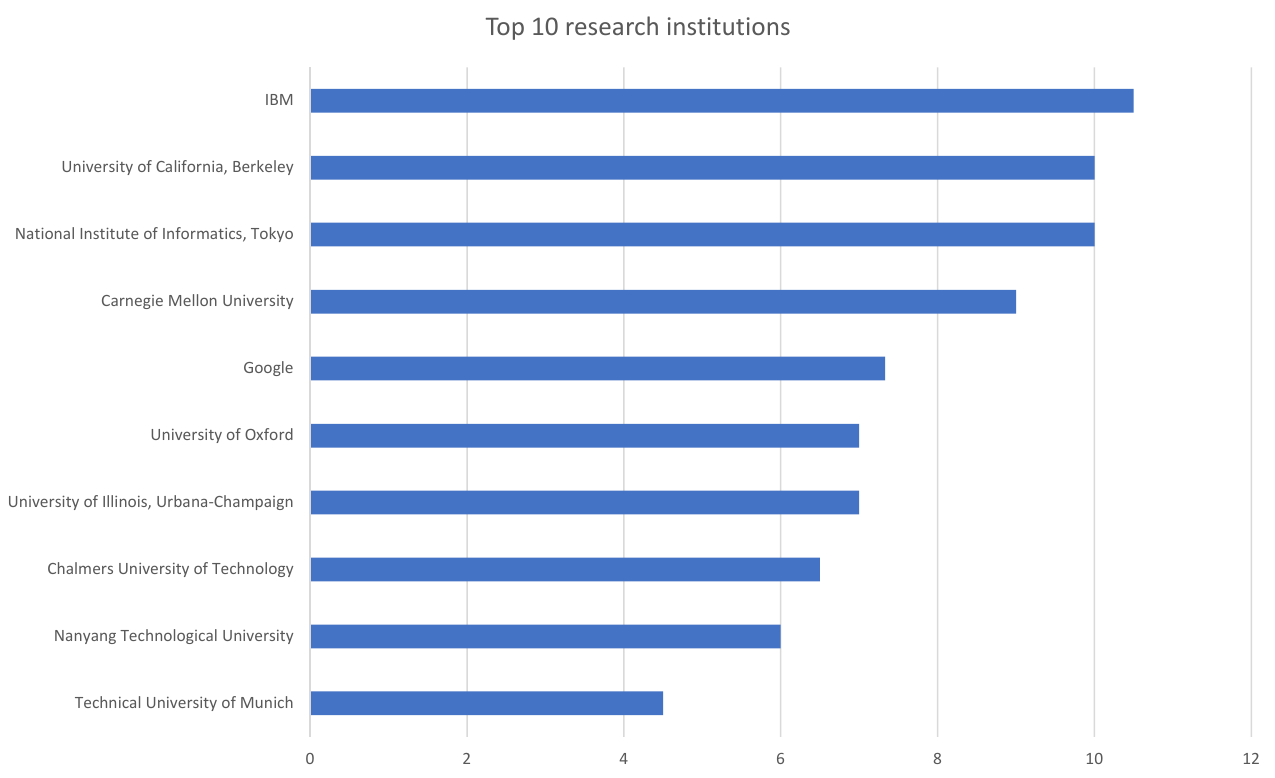}
    \caption{Number of publications by first author's affiliation (top 10 institutions).} \label{fig:Figure_4_4b}
\end{figure}

\textbf{Top research institutions.} We identified the top 10 research institutions based on the affiliation of the first author of the papers (see Figure~\ref{fig:Figure_4_4b}). It is worth mentioning that 23 authors had two affiliations and, in these cases, each of their affiliations counted as 1/2. Analogously, one author had three affiliations, and each of his affiliations counted as 1/3. As shown, the leading research institution is IBM, followed by National Institute of Informatics of Tokyo, University of California at Berkeley, Carnegie Mellon University, and Google.

\subsection{Characteristics of the primary studies}
\label{sec:characts}

From the 248 studies analyzed, 156 were empirical studies. We classified these empirical studies according to the research method reported. Table~\ref{tab:Tab_4_1} provides the list of research methods, the criteria used to classify them, and the number of papers for each type of research method. Interestingly, the most common type of study is those who provided one (or more) case studies (37.8\%). We also notice that SE4AI has a significant number of benchmarks (22.4\%) which might be explained due to the data-driven nature of the field. 

To provide some context to the distribution of papers by research method in SE4AI, we may compare the obtained results with the bibliometric assessment of SE by Wong et al.~\cite{Wong2021}. As part of a series of bibliometric reports on SE, they analyzed SE papers from 2013 to 2020. It is worth noting that the classification of papers by research method in~\cite{Wong2021} relies on a mapping of terms found in the titles of the primary studies only. Hence, a formal comparison with our findings in quantitative terms cannot be performed directly. Nevertheless, their results are quite consistent with ours. From 11,500 papers analysed in~\cite{Wong2021}, the most common research methods in SE are (as reported in the titles of primary studies): case studies (315 papers), experiments (194), literature reviews (179), and surveys (138), among others (e.g. simulation, theory, systematic mapping).

This distribution of papers by research method is in line with our findings for SE4AI, where the top research methods in SE4AI are: case studies (59 papers), benchmarks (35), experiments (23) and surveys (14). The only differences are that we do not include "literature reviews" (which is part of our Exclusion Criteria - EC4) and that we found a significant number of benchmarks. However, we must acknowledge that none of the papers classified as benchmarks in our study included the term benchmark in the title.

\begin{table}
\caption{Number of papers per type of research method.}
\begin{tabular}{|l|l|l|l|}
\hline
\textbf{Research method} & \textbf{Classification criteria} & \multicolumn{1}{c|}{\textbf{\#papers}} & \multicolumn{1}{c|}{\textbf{\%}} \\ \hline
Case study & \begin{tabular}[c]{@{}l@{}}The study reports that it employs a case study or an exploratory\\  study where the researchers analyze and answer predefined\\ questions for a single or multiple cases.\end{tabular} & 59 & 37.8\% \\ \hline
Benchmark & \begin{tabular}[c]{@{}l@{}}A rigorous study that evaluates or compares one or more\\ algorithms in well established operational settings (with data \\ and variables).\end{tabular} & 35 & 22.4\% \\ \hline
Experiment & \begin{tabular}[c]{@{}l@{}}An empirical enquiry that investigates causal relations and \\ processes.\end{tabular} & 23 & 14.7\% \\ \hline
Survey & \begin{tabular}[c]{@{}l@{}}The study reports that it employs a survey through a\\ questionnaire, observation, or interview.\end{tabular} & 14 & 8.9\% \\ \hline
Mixed method & \begin{tabular}[c]{@{}l@{}}The study reports using several research methods or a mix of \\ research methods.\end{tabular} & 5 & 3.2\% \\ \hline
Controlled experiment & \begin{tabular}[c]{@{}l@{}}The study mentions that it employs an empirical enquiry that\\ manipulates one factor or variable of the studied setting.\end{tabular} & 4 & 2.5\% \\ \hline
Action Research & \begin{tabular}[c]{@{}l@{}}The study reports employing action research. That is, a research\\ idea is applied in practice and the results are evaluated (a \\ crossing between an experiment and a case study).\end{tabular} & 1 & 0.6\% \\ \hline
Other / Not stated & \begin{tabular}[c]{@{}l@{}}Other research methods, or the research method was not stated \\ in the empirical study.\end{tabular} & 15 & 9.6\% \\ \hline
\end{tabular}
\label{tab:Tab_4_1}
\end{table}

For the quality assessment, we applied Ivarsson and Gorscheck's rigor and relevance quality assessment model~\cite{ivarsson2011method}. From the different metrics proposed in this model, we report here only those in which we feel confident that they have been measured in the most objective and consistent manner by all the different authors of this study. In other words, we found that, despite several efforts, evaluating some metrics is prone to subjective interpretation (e.g. measuring if a study design is described with enough rigor) and we did not feel confident that the results of those metrics were consistent enough to provide a reliable analysis. The metrics that we have finally incorporated in the report of the results are: the realism of the study environment, the scale, and whether they included threats to validity.    

\subsubsection{Evaluation of the realism of the study environment}

\begin{figure}
  \centering
    \includegraphics[width=0.85\textwidth]{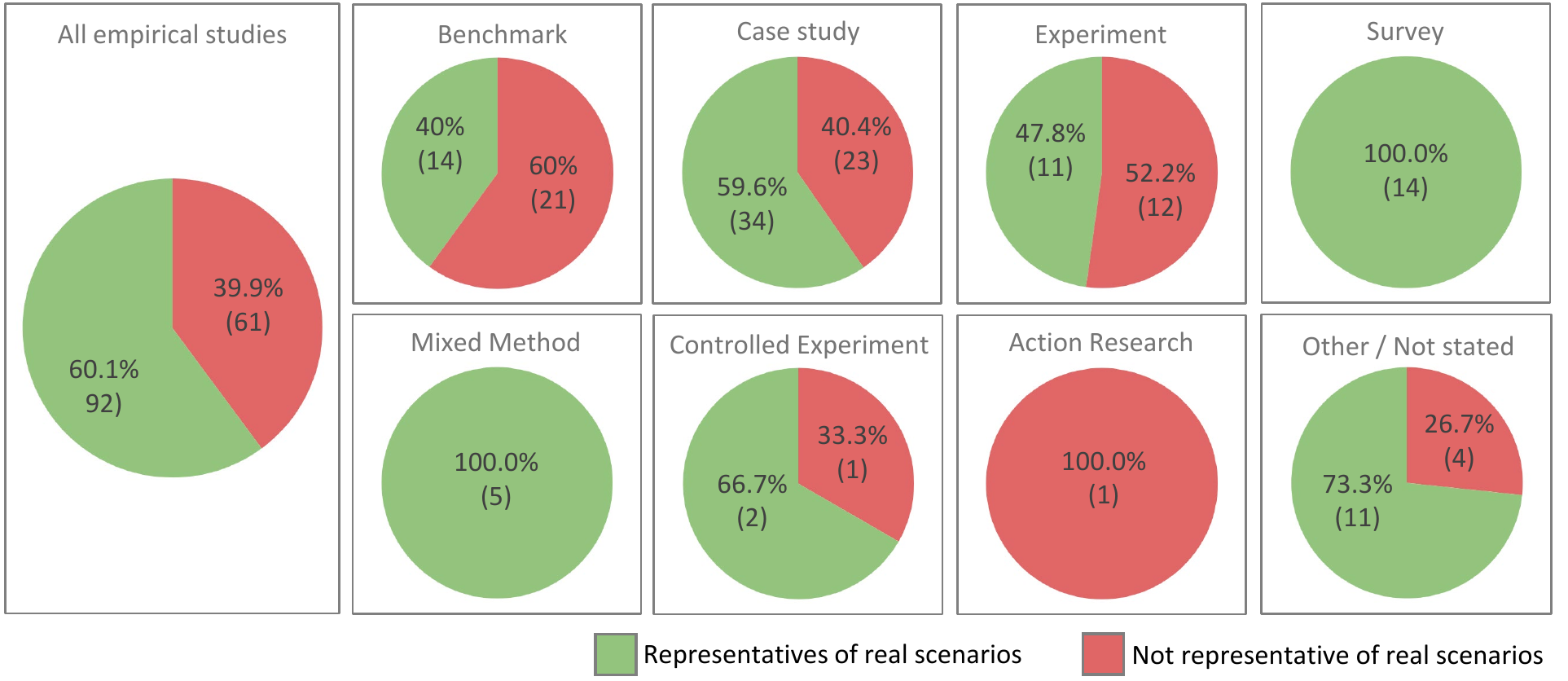}
    \caption{Subjects/Objects of study in empirical research studies.} \label{fig:Figure_4_5}
\end{figure}

Figure~\ref{fig:Figure_4_5} presents the evaluation of the realism of the study environment in terms of the subject/object used in the empirical study. In 60.1\% of empirical research studies, the subjects and/or objects used in the evaluation are representatives of industrial professionals and industry systems or real data sets). Conversely, 39.9\% of papers used scenarios based in students and/or simulated data settings.

Analyzing these results by type of study, we see that all surveys and mixed methods address representative subjects/objects. Although this is not surprising for surveys, as they generally approach subjects with the required background, it is quite interesting to see that also mixed methods have such a high degree of real subjects/objects (although, given the low number of studies of this type, this is not a firm conclusion and should be taken with care). In contrast, papers using more non-representative subjects/objects are benchmarks (60\%) and experiments (52.2\%) (we refrain from drawing conclusions on action research, as in this type of study n=1). These results are not surprising considering the nature of this type of studies (especially in the case of experiments).  

\begin{figure}
  \centering
    \includegraphics[width=0.85\textwidth]{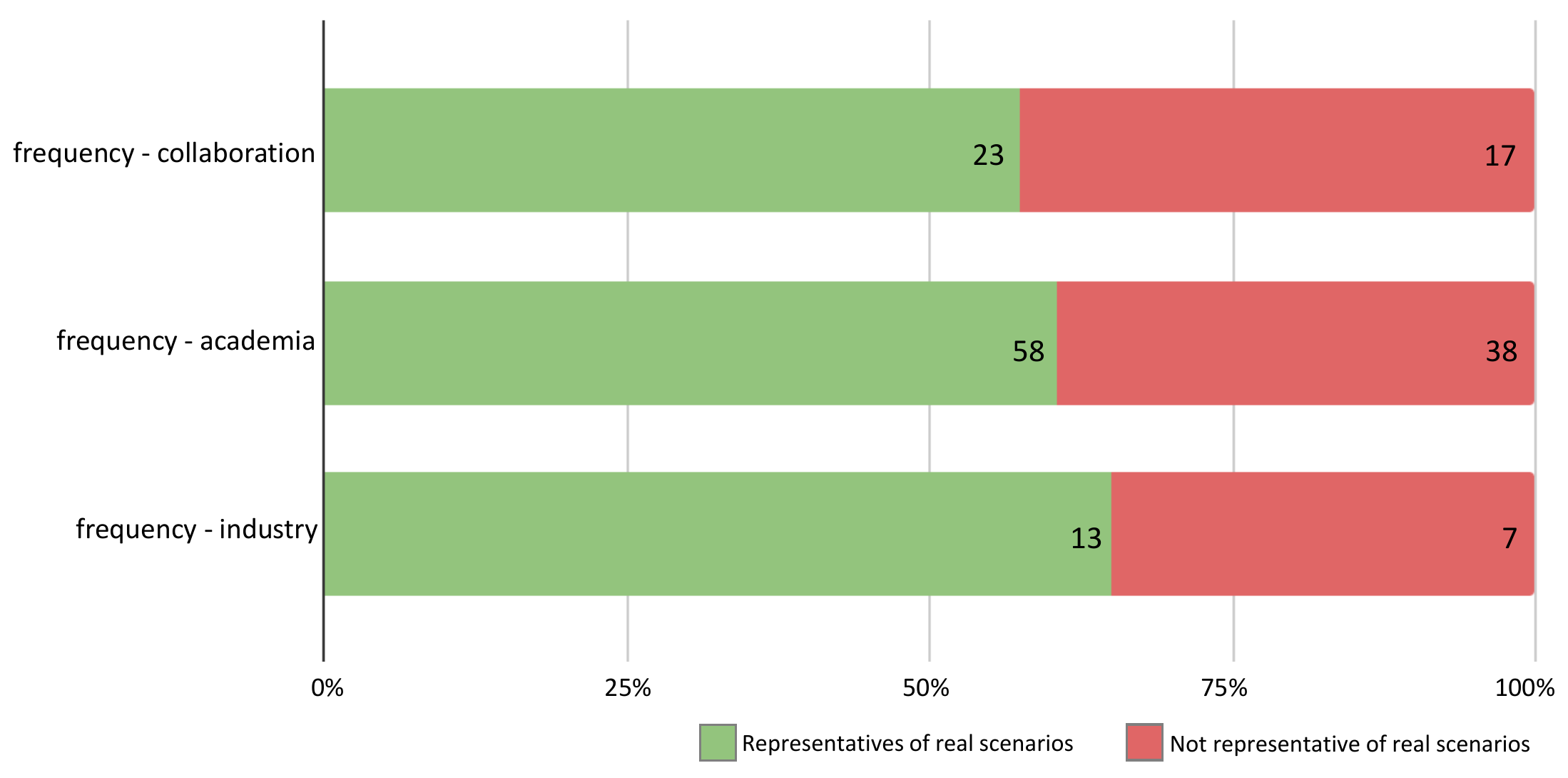}
    \caption{Subjects/Objects of study in empirical research studies by authors’ affiliation.} \label{fig:Figure_4_6}
\end{figure}

Considering the author’s affiliations (see Figure~\ref{fig:Figure_4_6}), we observe that papers written by authors from the industry have a slightly higher percentage of using real scenarios compared to collaboration or purely academic papers. 

\subsubsection{Scale of the application used in the evaluation}

\begin{figure}
  \centering
    \includegraphics[width=0.85\textwidth]{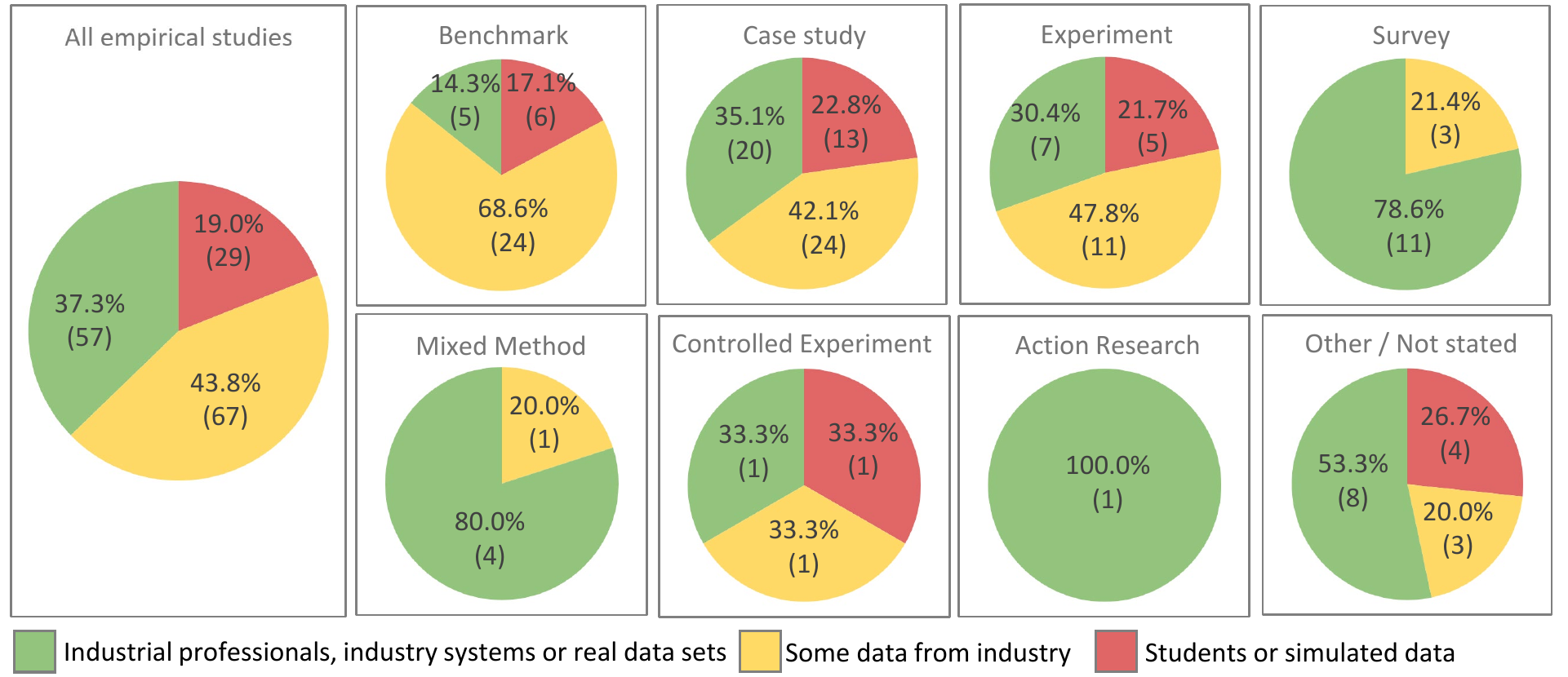}
    \caption{Scale of the application used in the evaluation.} \label{fig:Figure_4_7}
\end{figure}

Figure~\ref{fig:Figure_4_7} presents the scale of the application used in the evaluation. In 37.3\% of the empirical studies, the scale of the application used was of industrial scale. In 43.8\% of cases, the evaluation included some reference to the industry (e.g. data from a company) and in 19.0\% the scale was of laboratory data (i.e. toy examples). 

If we analyze the results by type of study, we observe that surveys and mixed methods present mostly evaluations of realistic size (for mixed methods, this is not a firm conclusion due to the low number of studies of this type, and again, we refrain from drawing conclusions on action research). Benchmarks, case studies and experiments usually use some data from the industry (68.6\%, 42.1\% and 47.8\%, respectively) but those with a scale of realistic size are very low (14.3\%, 35.1\% and 30.4\%). The rest of types of study do not have enough studies to present any valid conclusion. 

\begin{figure}
  \centering
    \includegraphics[width=0.85\textwidth]{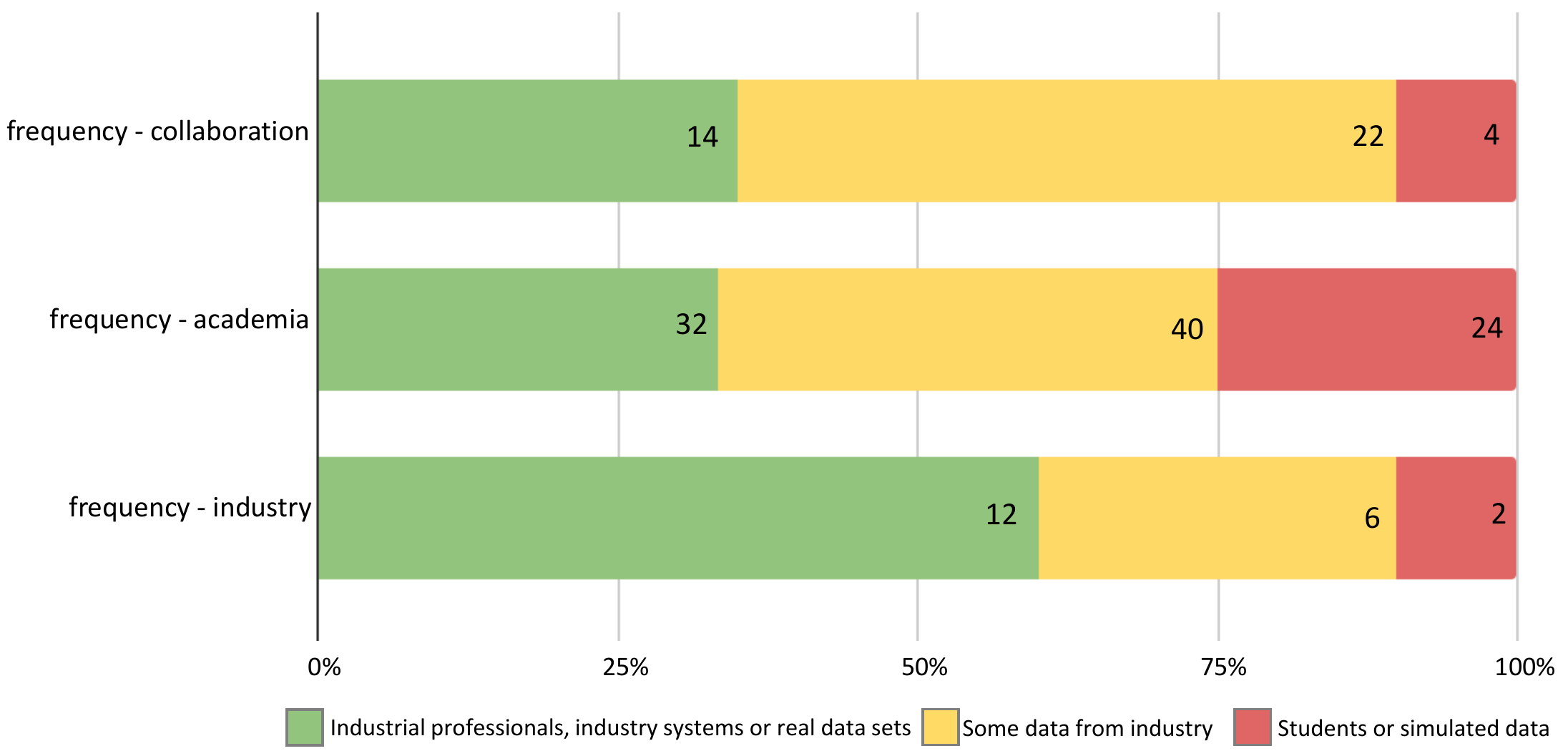}
    \caption{Scale of the application used in the evaluation by authors’ affiliation.} \label{fig:Figure_4_8}
\end{figure}

Considering the authors’ affiliation (see Figure~\ref{fig:Figure_4_8}), we observe that most of the papers written by industry authors used an application of realistic size, whereas papers from both collaboration and purely academic authors had most of the papers including some data from the industry but without applying it into a realistic size environment. Finally, academic papers were the ones that had most of the papers using toy examples.

\subsubsection{Threats to validity in the primary studies}

\begin{figure}
  \centering
    \includegraphics[width=0.85\textwidth]{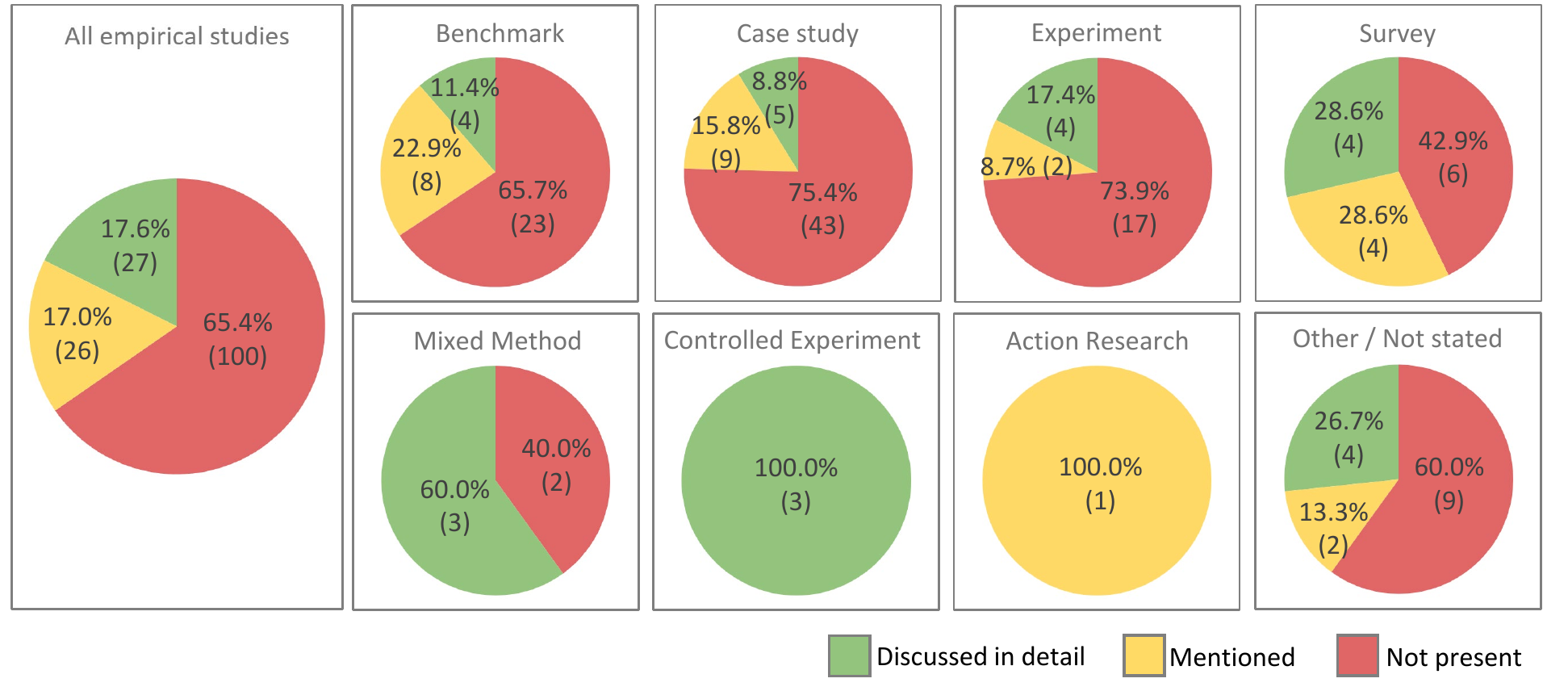}
    \caption{Threats to validity in empirical research studies.} \label{fig:Figure_4_9}
\end{figure}

Strikingly, 65.4\% of the empirical research studies do not provide any threat to validity. Only 17.6\% of these studies have the validity of the evaluation discussed in detail, and the remaining 17.0\% just briefly mention them (see Figure~\ref{fig:Figure_4_9}). 

If we analyze these results by type of empirical study, we notice that research papers presenting case studies, experiments and benchmarks are the ones where threats to validity are mostly ignored. Most of these research papers do not even mention threats to validity, ignoring them in 75.4\%, 73.9\% and 65.7\% of the cases, respectively. In contrast, those discussing more in detail the threats to validity are controlled experiments and mixed methods (again, this conclusion should be taken with care due to the low number of papers of these types).

\begin{figure}
  \centering
    \includegraphics[width=0.85\textwidth]{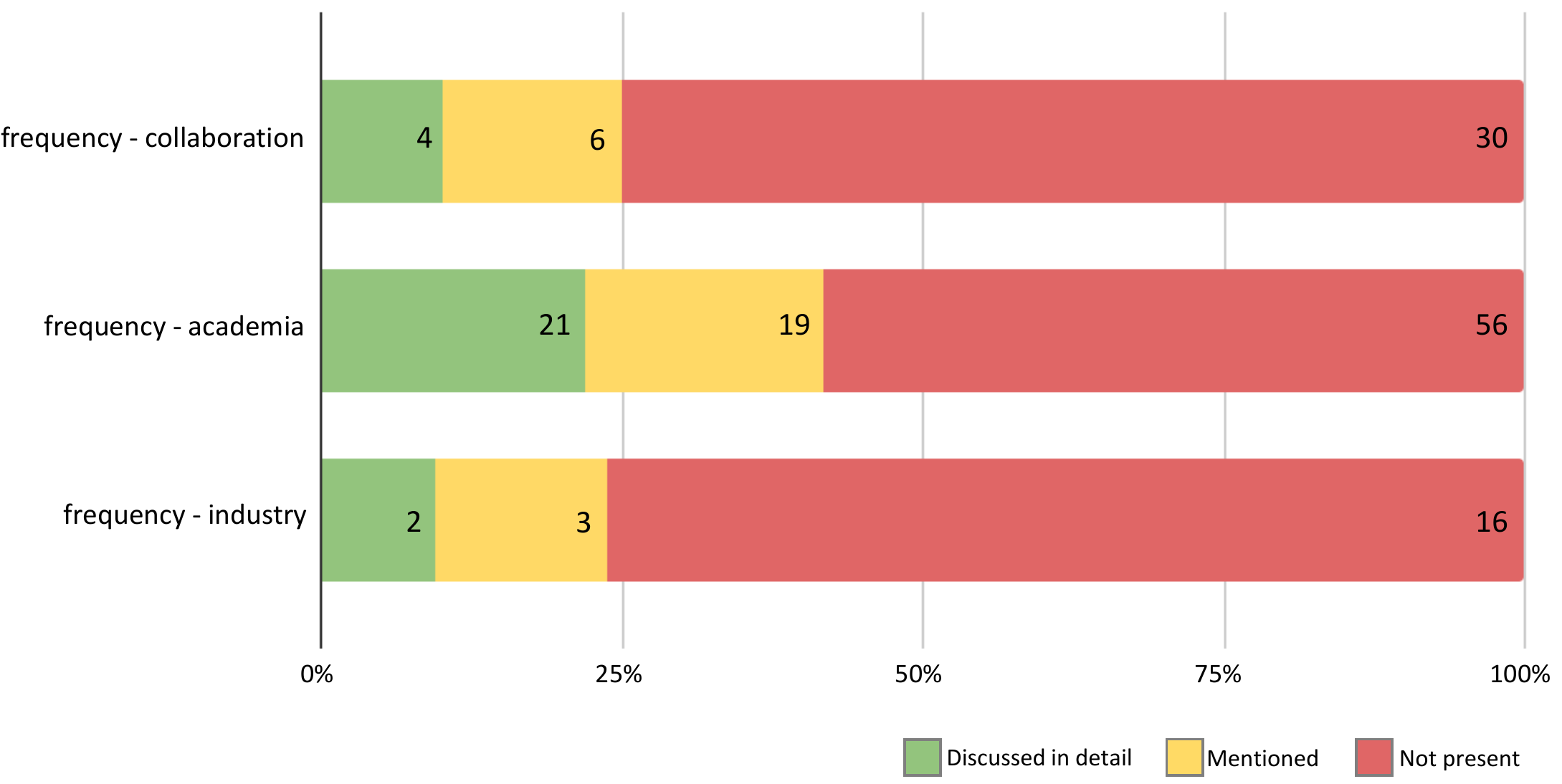}
    \caption{Threats to validity in empirical research studies by authors’ affiliation.} \label{fig:Figure_4_10}
\end{figure}

Analyzing the results by authors’ affiliation (see Figure~\ref{fig:Figure_4_10}), we observe that the vast majority of papers from collaborations and industry do not discuss threats to validity. But surprisingly, most of the academic papers do not discuss threats to validity either, even though the frequency of papers that discuss threats to validity is higher than its counterparts.

\subsection{Discussion}

Below, we report the main observations and take-away messages for this RQ:

\textbf{Observation 1.1: SE4AI is an emerging research area.} Not only the growing annual publication trend supports this observation, but also the distribution in type of venues, with the importance of arXiv and the low percentage of publications in the form of journal papers (26 papers, i.e. 10.5\%; with 23 of them published in the last three years). We may expect that a number of these arXiv publications will become archival publications in journals, contributing to the gradual consolidation of the field. 

\textbf{Observation 1.2: Literature reviews in SE4AI need to consider arXiv.} As a follow-up of the previous observation, we may affirm that literature reviews in the SE4AI area cannot be limited to automatic searches in typical digital libraries (Scopus, Web of Science, or publisher digital libraries as ACM DL or IEEE Xplore) because arXiv papers will not be found. Either arXiv papers are manually added to the results, or snowballing is used, as we have done in our study.

\textbf{Observation 1.3: Research in SE4AI involves more industry authors than usual.} Our study has identified 101 industry and collaboration papers, representing 40.7\% over the total. We have compared these numbers with other recent literature reviews on topics that can be considered of practical importance, in which we find lower percentages, e.g. 25.6\% in Management of quality requirements in agile and rapid software development~\cite{behutiye2020management} or 28.7\% in open-source software ecosystems~\cite{franco2017open}. This observation is important to argue for practical applicability of the findings reported in the primary studies found in our literature review. Major players in the industry are big companies like Microsoft, IBM or Google.

\textbf{Observation 1.4: Industry involvement is especially significant in Europe and North America.} Looking in more detail the results of industry involvement, the percentage grows significantly in North America (45.5\%) and  Europe (48.6\%) compared to Asia (21.8\%). This difference becomes more apparent if we compare the two countries with the highest numbers of studies, the USA (46.5\%) and China (8.7\%), showing two different approaches to research.

\textbf{Observation 1.5: Industry involvement slightly improves the realism of case studies, and significantly improves its scale.} Industry papers have just a slightly higher percentage of realistic scenarios compared to collaboration or academic papers. In contrast, we observe that the authors’ background affects more significantly the scale of the evaluation. In this regard, industry papers use bigger scales in the evaluations, with approximately, twice as much as collaboration or academic papers. Conversely, academic papers use more toy examples compared to its counterparts, with approximately three times as much as collaboration or industry papers.

\textbf{Observation 1.6: Threats to validity are mostly ignored, even for papers from academic authors.} 
As observed, most empirical studies do not discuss threats to validity. Papers from academic authors have a higher frequency of papers discussing threats to validity compared to collaborations or industry papers, but they are still a minority, and most of the academic papers ignore threats to validity, which compromise the quality of the research. This may be caused by the number of arXiv and workshop papers, which tend to discuss fewer threats to validity than journal or conference papers.

\section{RQ2: What are the characteristics of AI-based systems?}\label{sec:results-rq2}

This section respectively discusses the terminology used on the primary studies as well as the dimensions in which we have classified them, and the key quality attribute goals of AI-based systems.

\subsection{What is an AI-based system?}

We found a large variety of terms used in the primary studies. In Table~\ref{tab:Tab_5_1}, there is an overview of all terms used more than twice to discuss the type of system investigated in the corresponding primary study. We observed a mix of very general terms (such as "machine learning" or "AI technologies"), specific AI technologies (such as "deep neural networks" or "ML libraries") and AI application domains (such as "robotics system" or "automotive system").
We furthermore noticed that AI seems to be always used in terms of learning components and not including rule-based expert systems. This corresponds to the new wave of AI associated with learning from data.

\begin{table}
\caption{Terms used in the primary studies to refer to AI-based systems with "intelligent" components.}
\begin{tabularx}{\textwidth}{|l|X|}
\hline
\textbf{Count} & \textbf{Terms (comma-separated)} \\ \hline
43 & Machine learning \\ \hline
40 & ML system \\ \hline
28 & Deep neural networks \\ \hline
27 & ML algorithms \\ \hline
23 & ML models \\ \hline
21 & Autonomous vehicle \\ \hline
19 & Autonomous systems, ML components \\ \hline
18 & AI systems \\ \hline
17 & AI, Neural network \\ \hline
16 & Deep learning systems, ML application \\ \hline
15 & ML techniques \\ \hline
14 & Autonomous driving system \\ \hline
11 & Cyber-physical systems with Machine Learning components (CPSML) \\ \hline
10 & ML software, ML-based system \\ \hline
9 & AI-based system \\ \hline
8 & DNN-based software \\ \hline
7 & Deep learning, Reinforcement Learning (RL) \\ \hline
6 & AI software, Machine learning classifiers, ML program \\ \hline
5 & Artificial Neural Networks, Classifier, Intelligent systems \\ \hline
4 & AI components, AI model, DNN model, ML methods, ML pipeline \\ \hline
3 & AI applications \\ \hline
\end{tabularx}
\label{tab:Tab_5_1}
\end{table}

In the further inductive coding of the terms and study objects, we distilled three dimensions that can be used to classify the contributions of primary studies about AI-based systems. 

The first dimension is the \textbf{Scope} of the system under analysis. In particular, the scope refers to the question: \textit{How is AI implemented inside the system?} AI can either be one component in a system ("component"), dominating the entire system ("system"), implement one or more particular algorithms ("algorithm", such as DNN), or providing a pipeline or infrastructure ("infrastructure", e.g., PyTorch).

The second dimension is the \textbf{Application Domain}. Many studies do not concern themselves with what the AI will be used for exactly but study, for example, DNNs in general. There are, however, also several studies focusing on concrete applications of AI. The most frequent of those is the domain of autonomous vehicles that encompasses a part of the autonomous systems and the autonomous/automated driving systems from Table~\ref{tab:Tab_5_1}. An application domain can also be more generic, such as ML frameworks.

The third dimension is the \textbf{Technologies of AI} under consideration. AI technologies can be, for example, ML in general or DL methods. This is usually stated in some way, and we also think it is important because it can make a huge difference in how generalizable the results of a primary study are. For instance, test approaches for AI components that use random forests might not be useful for AI components built on DNNs.

\begin{figure}
  \centering
    \includegraphics[width=0.7\textwidth]{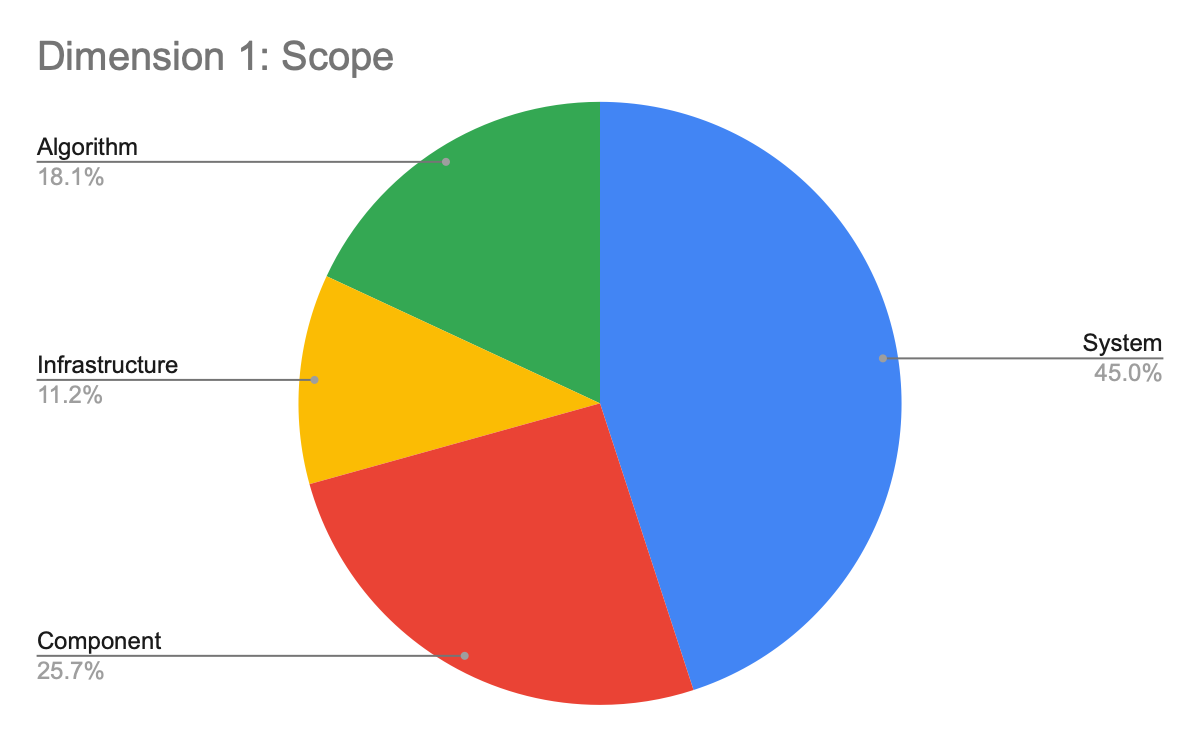}
    \caption{Scope of the research in the primary studies.} \label{fig:Figure_5_2}
\end{figure}

We then used this structure to code all our primary studies (not only those that provided definitions) to get insights into what exactly they investigated. We found that almost half of the primary studies look at SE4AI at the system level (see Figure \ref{fig:Figure_5_2}). This means that they investigate complete systems such as autonomous cars regarding their AI aspects. A quarter of the primary studies focuses on the AI components directly. An example would be the image recognition component in an autonomous car. The remaining primary studies either investigate or propose methods for specific algorithms (such as DL) or for AI infrastructure (such as TensorFlow) without considering specific applications.

\begin{figure}
  \centering
    \includegraphics[width=\textwidth]{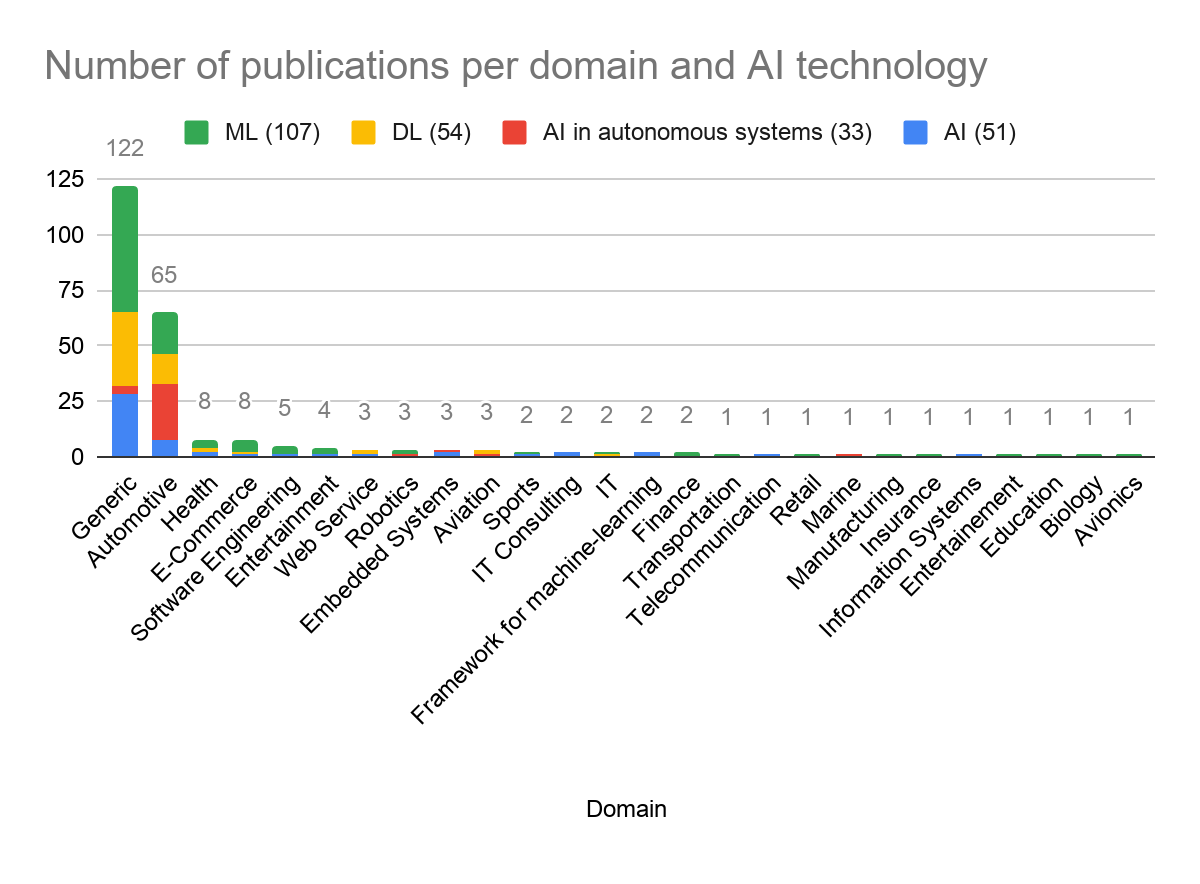}
    \caption{Domain and AI technology investigated in the primary studies.} \label{fig:Figure_5_3}
\end{figure}

In Figure \ref{fig:Figure_5_3}, we show the number of publications in different domains structured by the AI technology investigated. The only dominant application domain is automotive with its hype on autonomous cars. More than a quarter of the primary studies aim at this domain. Example systems are the pedestrian detection system in autonomous driving at Bosch \cite{Gauerhof2018} or an automated emergency braking system at IEE S.A. (Luxembourg) \cite{Abdessalem2018}. Almost half of the primary studies look at AI in a generic way without considering any application domain. The further many domains that we found only make up between 0.8\% and 3.3\% of the primary studies. Some domains also overlap, as there are embedded systems in automotive or aviation. Examples for these further domains include WeChat's NMT system for automatic machine translations \cite{Zheng2019}, the pin recommender system at Pinterest \cite{Liu2017} or the CognIA chatbot for financial advice from IBM \cite{Vasconcelos2017}.

\subsection{What are the key quality attribute goals for AI-based systems?}

\begin{figure}[H]
    \centering
    \includegraphics[width=0.85\textwidth]{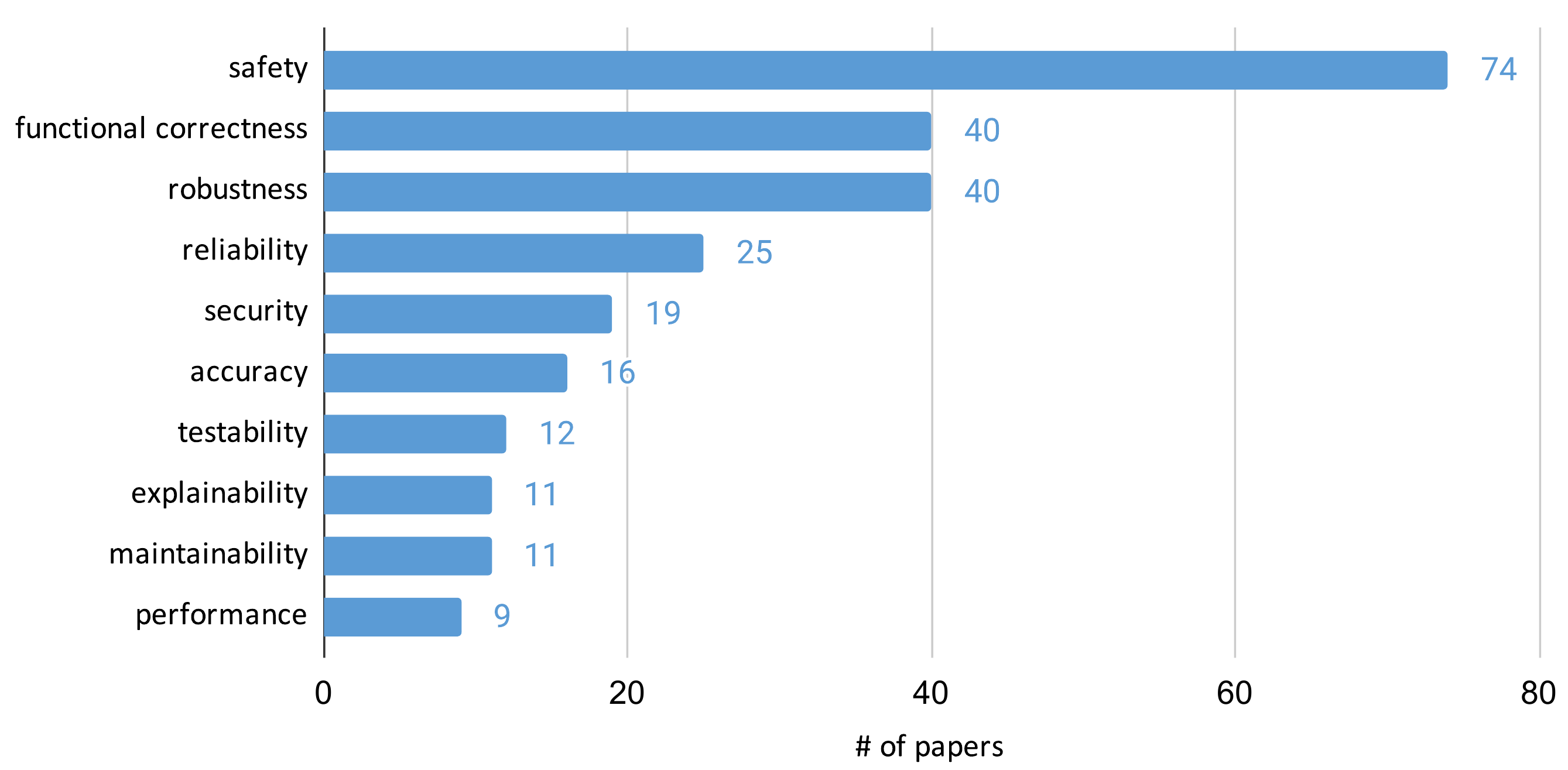}
    \caption{Most frequent quality goals of AI-based software systems.}
    \label{fig:Figure_5_4}
\end{figure}

Many of the primary studies focused on one or several software product or process qualities, e.g. by proposing an approach that improves a certain quality attribute or by analyzing the context of a certain quality for AI-based systems. We therefore extracted the quality goals per study (0\ldots n) and could assign at least one goal to 190 out of 248 studies. These study goals were then harmonized for consistent terminology, until we ended up with 40 different terms (see Figure~\ref{fig:Figure_5_4} for the most frequent ones). In total, these terms were mentioned 378 times, i.e. each study was linked to an average of 1.5 quality attribute goals. We then analyzed this mapping to identify trends and generalizations. For the analysis, we first identified the level of abstraction per goal (vertical analysis) and then formed thematic clusters of semantically related goals (horizontal analysis).

Regarding the vertical analysis, the identified publications discuss 40 quality attributes at various \textit{levels of abstraction}, which we classified with the three labels \texttt{low} (very specialized low-level attributes, e.g. sub-QAs in ISO 25010), \texttt{medium} (top-level attributes in ISO 25010 or similar granularity), and \texttt{high} (very abstract QAs or aspects encompassing a broader range of qualities). Exactly half of them (20) were classified as \texttt{low}, 13 as \texttt{medium}, and 7 as \texttt{high}. This indicates that the majority of approaches aims to improve quality attributes at the abstraction level specified in ISO 25010 or the level below, with hardly any studies targeting more abstract or broad-range qualities. The most frequently discussed quality attributes include \textit{functional correctness} (40 mentions) at the lowest level, \textit{safety} (74 mentions) at the medium level, and \textit{trust} (8 mentions) at the highest level.

\begin{figure}[H]
    \centering
    \includegraphics[width=\textwidth]{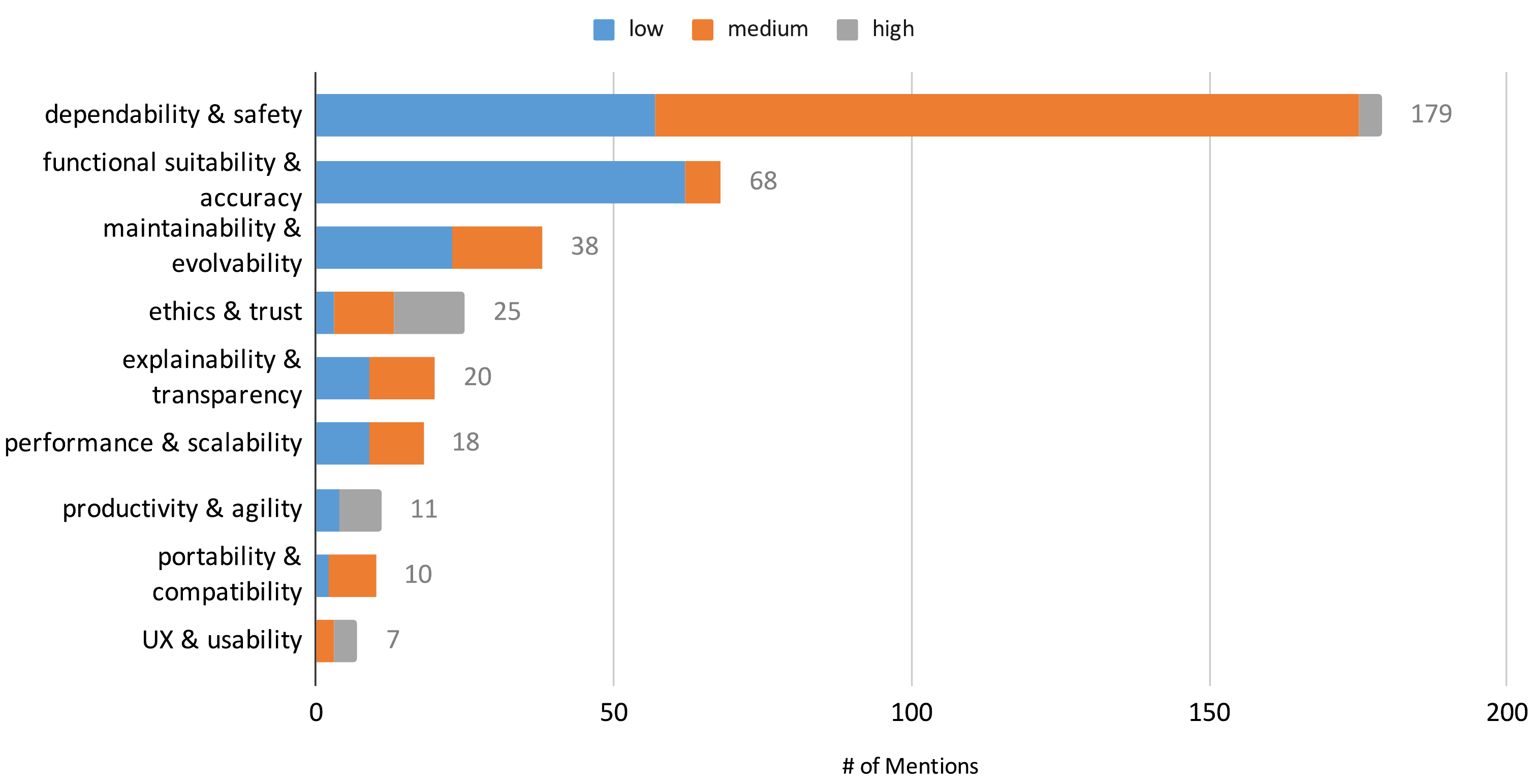}
    \caption{Frequency of quality goals per thematic cluster and level of abstraction.} \label{fig:Figure_5_5}
\end{figure}

For the horizontal analysis via thematic clustering, we identified a total of nine clusters (see Figure~\ref{fig:Figure_5_5}).  Except for the goal for general quality attributes, each of the 40 goals and their associated studies are assigned to exactly one of these clusters. By far the most prominent cluster is \textit{dependability \& safety}, which accounts for a total of 179 mentions. It includes four of the five most frequently mentioned individual QA goals, namely \textit{safety} (74), \textit{robustness} (40), \textit{reliability} (25), and \textit{security} (19). With 68 mentions, \textit{functional suitability \& accuracy} is the second-largest cluster, followed by \textit{maintainability \& evolvability} with 38 mentions. The remaining six clusters are smaller (7 - 25 mentions). Overall, we identified a strong focus on qualities related to safety (especially in the context of smart cyber-physical systems like autonomous vehicles) and correctness \& accuracy, which was analogous to the many studies on AI software testing.

\subsection{Discussion}

\textbf{Observation 2.1: AI is commonly associated with DL and considered as part of a complex software system.} We propose that any paper that fits into our inclusion criteria discussing the application of SE to AI-based systems should make explicit what exactly they consider on our identified dimensions. This would be a first step in making the scope and limitations of primary studies clearer. 

\textbf{Observation 2.2: Many various synonymous terms are used to denote a system or component that uses some kind of AI or ML.} Commonly used synonyms used to refer to software systems, which use AI technologies include AI-based system, AI-enabled system, AI-infused system, AI software, ML solution or DL system. We propose to use the following definitions to guide the selection of terms for future studies:
\begin{itemize}
  \item \textbf{AI component:} A part of a system. The component uses AI to some extent. Examples range from a component whose behaviour  depends to some extent on some embedded AI code, or an AI library as a special AI component that provides a concrete implementation of AI algorithms.
  \item \textbf{AI-based system:} A system consisting of various software and potentially other components, out of which at least one is an AI component.
  %\item \textbf{AI library:} A special AI component that provides a concrete implementation of AI algorithms.
\end{itemize}

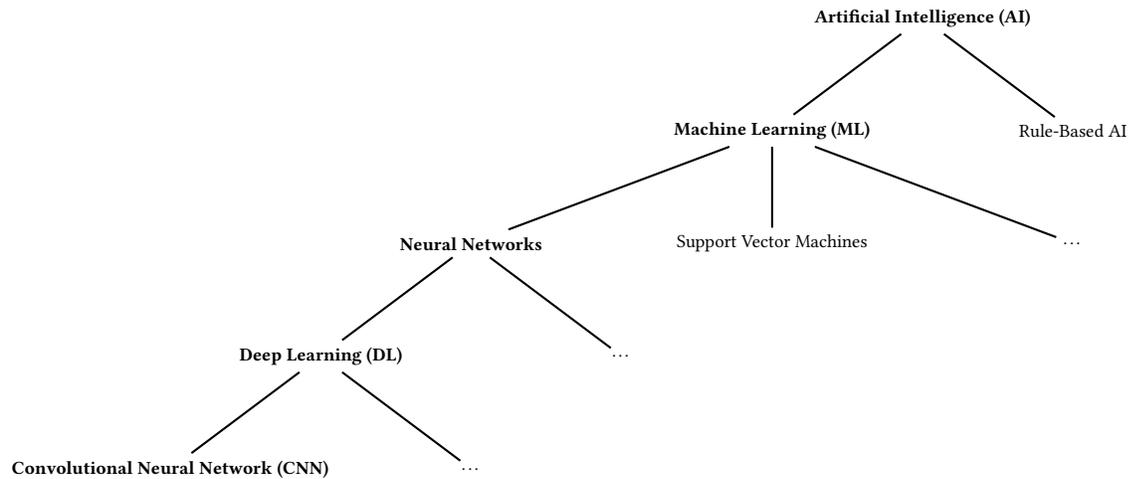
\begin{figure}
\begin{tikzpicture}
[sibling distance=4cm,-,thick]
\footnotesize
\node {\textbf{Artificial Intelligence (AI)}}
  child {node {\textbf{Machine Learning (ML)}}
    [sibling distance=4cm]
    child {node {\textbf{Neural Networks}}
      child {node {\textbf{Deep Learning (DL)}}
        [sibling distance=4cm]
        child {node{\textbf{Convolutional Neural Network (CNN)}}}
        child {node {\ldots}}
        }
      child {node {\ldots}}
    }
    child {node {Support Vector Machines}}
    child {node {\ldots}}
  }
  child {node {Rule-Based AI}
  };
\end{tikzpicture}
    \caption{Example taxonomic classification for paper \cite{Burton2017}. \label{fig:Figure_5_6}}
\end{figure}

\textbf{Observation 2.3: There exists diversity in the terminology to refer to AI-based systems, making unclear what is the object of the research.} We propose that SE4AI papers should use a taxonomy that makes clear what kind of AI-based system is investigated and how general the methods and approaches are supposed to be. We use as an example the paper by Burton et al. \cite{Burton2017}. They use the term ``machine learning'' in the title, but then focus on Convolutional Neural Networks. We depicted a corresponding taxonomy in Figure \ref{fig:Figure_5_6}. On the top level, we keep as close as possible to established discussions of the terms. For that, ML is commonly seen as a part of AI (\cite{Kuehl2019machine}). Hence, an ML-based system is also an AI-based system. Using the DARPA terminology (\url{https://www.darpa.mil/news-events/2018-07-20a}), in the first wave of AI, there were mostly rule-based systems. The second wave added statistical learning. We see only papers about AI in at least the second wave sense in our primary studies. Similarly, on the next level, in ML most primary studies investigate neural networks, mostly DL. So, also for Burton et al., we would go further and also refine it to the next level ``Convolutional Neural Network'' as a specific type of DL. Using such a taxonomic classification and explicitly mapping contributions to their respective levels would make the article clearer. We suggest that such a taxonomic classification would be useful for all SE4AI papers. %Please note that also adding the distinction of supervised, unsupervised, semi-supervised and reinforcement learning could be useful here, but it is a separate dimension from the taxonomy shown in Figure \ref{fig:Figure_5_6}.

\textbf{Observation 2.4: Most of the terms used are not defined explicitly.} Most commonly defined terms include "deep learning system" (5), "ML system" (5), "AI" (4), and "machine learning" (4). Besides using a taxonomy, we propose explicitly defining the terms used.

\textbf{Observation 2.5: Most primary studies focus on software systems.} In the analyzed studies, AI is not just one of its many components of a software system, but typically constitutes its dominating part.

\textbf{Observation 2.6: The most studied properties of AI-based systems are dependability and safety.} Overall, we identified a strong focus on qualities related to safety (especially in the context of smart cyber-physical systems like autonomous vehicles) and correctness \& accuracy. However, there are research gaps for less studied properties, such as usability, portability or particularly important in the context of trustworthiness and the understandability aspect. More importantly, inherent and critical quality characteristics in AI-based systems such as explainability and transparency require the attention of researchers to assure the high maturity level required by industry.

\textbf{Observation 2.7: The use of ML and DL has only been extensively used in AI-based systems of the automotive domain, and to a much lesser extent in healthcare and e-commerce.} We believe that the increasing and successful application of ML and DL (e.g., image-, language- classification, and object recognition) in the automotive domain shall inspire researchers and practitioners to explore and apply it in other domains. Based on the primary studies, many domains have received little or no attention (see Figure \ref{fig:Figure_5_5}).

\section{RQ3: Which SE approaches for AI-based systems have been reported in the scientific literature?}
\label{sec:results-rq3}
\begin{figure}
    \centering
    \includegraphics[width=\textwidth]{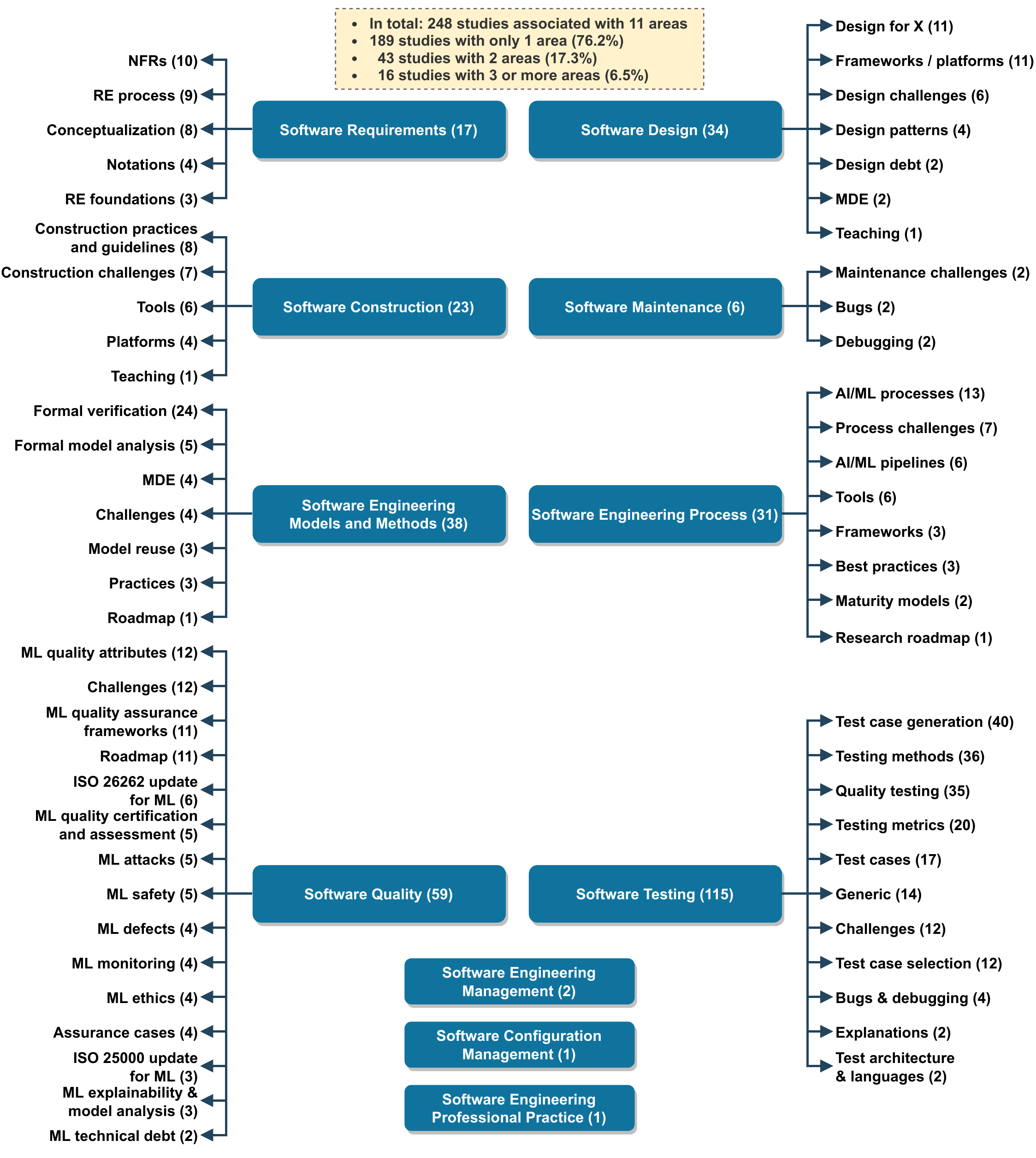}
    \caption{The 248 primary studies classified into 11 SWEBOK Knowledge Areas based on their SE contributions for AI-based systems.}
    \label{fig:Figure_6_1}
\end{figure}

We classified the 248 primary studies in 11 SWEBOK areas (see Figure~\ref{fig:Figure_6_1}).
Primary studies were thematically associated with at least one SWEBOK area based on their research directions and contributions.
For eight areas, we derived subcategories to organize the research further.
We did not do this for areas in which there were two or less primary studies.
In the following subsections, we provide a detailed overview of SE approaches for AI-based systems in each SWEBOK area and illustrate them with exemplary papers.

\begin{figure}
    \centering
    \includegraphics[width=\textwidth]{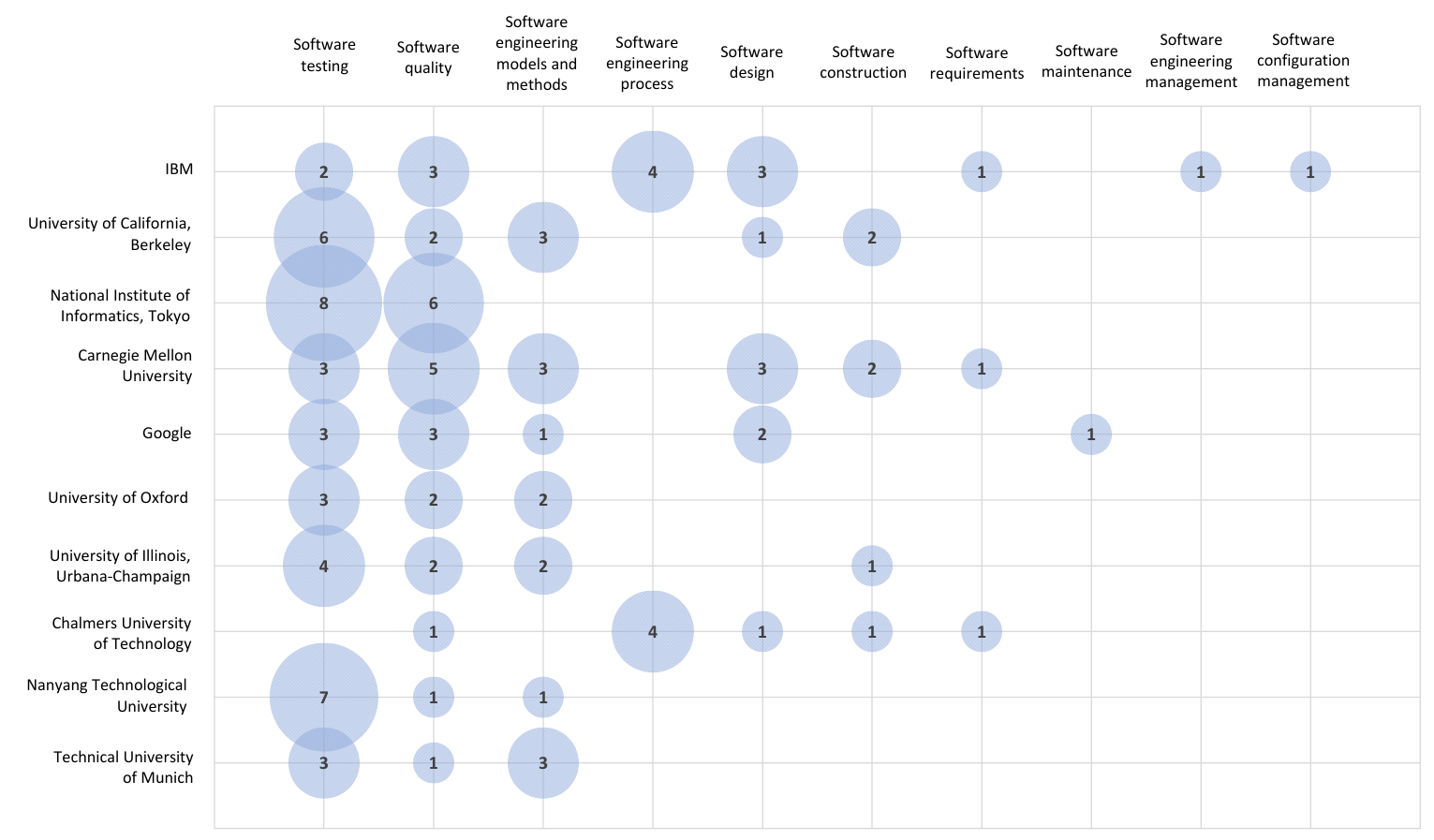}
    \caption{SWEBOK areas addressed by the leading institutions in SE4AI.}
    \label{fig:Figure_6_2}
\end{figure}

We also analysed the SWEBOK areas addressed by the leading research institutions in SE4AI (see Figure~\ref{fig:Figure_6_2}). In terms of research topic, we find a two-fold situation. Three institutions (University of California, Berkeley; National Institute of Informatics, Tokio; and Nanyang Technological University) are focused on one particular SWEBOK Knowledge Area, namely Software Testing, while Chalmers University of Technology conducts research mainly in the Software Engineering Process area. The rest of institutions cover a wider variety of topics, and particularly IBM and Carnegie Mellon University have published research related to 7 and 6 SWEBOK Knowledge Areas, respectively.

\subsection{Software requirements (17 studies)}
For \textit{software requirements}, we identified 17 studies. Based on our thematic analysis, we derived five subcategories, with most papers belonging to several categories. 

Nine papers address the Requirements Engineering (RE) \textit{process} for AI-based systems. Vogelsang and Borg are the only ones to cover the complete process, from elicitation to verification and validation \cite{Vogelsang2019}. They summarized important characteristics of the RE process for ML systems, such as detecting data anomalies or algorithmic discrimination. The rest of the primary studies concentrate on one particular RE activity, with \textit{specification} being the most popular one (six papers). Four of these papers focus on non-functional requirements (NFRs, see below for details). Further examples include methodological aids such as the notion of supplier’s declaration of conformity \cite{Arnold2018} or conceptual frameworks to improve the specification of requirements for explainable \cite{Sheh2018} or safe \cite{Banks2019} AI-based systems. The two general papers on specification focus on formal aspects related to ambiguity \cite{Rahimi2019} and the need to consider partial specifications for AI-based systems \cite{Salay2019}. Two other papers addressed requirements-driven \textit{derivation}, in both cases in automotive systems, but with different aims: while Burton et al. derive low-level requirements from safety goals \cite{Burton2017}, Tuncali et al. derive test cases from safety and performance requirements \cite{Tuncali2019}. The last paper focuses on the \textit{elicitation} of safety requirements in automotive systems using a risk-based approach \cite{Adedjouma2018}.

Four papers in the RE process category also propose a particular \textit{notation} to express requirements. The aforementioned two papers on derivation defined a concrete notation to make the derivation process less ambiguous, namely by using goals in the case of Burton et al. \cite{Burton2017} and temporal logic in Tuncali et al.’s approach \cite{Tuncali2019}. Adedjouma et al. also employ goals to represent safety risks and hazards \cite{Adedjouma2018}. The fourth paper expresses requirements as linear arithmetic constraints over real-valued variables to support satisfiability modulo theories (SMT) \cite{Leofante2016}. 

Eight papers did not provide concrete RE approaches, but were more \textit{conceptual} in nature: the authors tried to provide a foundation for AI software RE research by disseminating current \textit{practices} and/or \textit{challenges}, e.g. via the use of interviews \cite{Vogelsang2019} or surveys \cite{Zhang2020}. Many of these papers analyzed how RE for AI changed in comparison to traditional systems, and especially what issues currently prevent effective practices \cite{Belani2019, Horkoff2019, Kostova2020}. In addition to that, Otero and Peter also tried to provide research directions to address some of these challenges \cite{Otero2015}.

Finally, three very diverse papers discuss \textit{RE foundations} for AI-based systems. As mentioned above, Salay and Czarnecki introduced foundations for partial specifications as appropriate for specifying AI-based systems \cite{Salay2019}. Meanwhile, Otero and Peter propose a model of requirements for Big Data analytics software \cite{Otero2015}. Lastly, Arnold et al. incorporate traceability into their requirements specification approach \cite{Arnold2018}.

As an additional facet to the above subcategories, 10 of the 17 studies focused specifically on \textit{NFRs}. While both Horkoff \cite{Horkoff2019} and Kuwajima et al. \cite{Kuwajima2020} targeted NFRs in general, the other papers were concerned with one or a few specific NFRs, e.g. safety \cite{Adedjouma2018, Leofante2016, Banks2019, Salay2019} or model performance \cite{Burton2017, Arnold2018, Tuncali2019}. Arnold et al. also included security requirements in their FactSheets approach \cite{Arnold2018}, whereas Sheh and Monteath were the only ones to study the nature of requirements for explainable AI \cite{Sheh2018}.

\begin{tcolorbox}
    Key takeaways for software requirements:
    \begin{itemize}
        \item A lot of focus on NFRs for AI (10 of 17 studies), especially new AI-specific quality attributes
        \item Several specification and notation approaches to deal with probabilistic results or ambiguity, e.g. partial specification
        \item Very few holistic views on the RE process for AI (only one study), focus on support for RE specification and derivation
    \end{itemize}
\end{tcolorbox}

\subsection{Software design (34 studies)}
In the area of \textit{software design}, we formed seven categories to group the 34 studies. One of the largest of these -- \textit{design for X} -- is concerned with design approaches or techniques to improve or assure one specific quality attribute in AI-based systems. It comprises 11 papers, of which the majority is concerned with \textit{safety} (seven papers). Most of these studies propose design strategies for AI systems in safety-critical domains where unsafe behaviour can have immense negative consequences, like the use of architectural components as a protection layer for safety in autonomous vehicles \cite{Molina2017}, design best practices for safety verification methods \cite{Bansal2019} or behavior-bounded assurance \cite{Sarathy2019} for autonomous aerial vehicles, safety design strategies (e.g., inherently safe design, safety reserves, safe fail, and procedural safeguards) for ML components \cite{Varshney2016, Kuwajima2020}, and specific safety strategies for cyber-physical systems \cite{Amodei2016, Varshney2017}. The remaining four design approaches are for \textit{reliability} \cite{Machida2019b}, \textit{user experience} \cite{Yang2017}, and for quality attributes related to ethical AI like \textit{fairness} \cite{Arnold2018, Kumar2020}.

Another large category is \textit{frameworks or platforms} (11 studies), ranging from a generic level of abstraction for end-to-end ML application development and deployment \cite{Bailis2017, Colomo2019, Kumar2020, Khalajzadeh2018, Hummer2019} to available tool support for managing reliable ML applications \cite{Baylor2017}, packaging and sharing ML models as reusable microservices \cite{Zhao2018b}, and the development and deployment of ML applications \cite{Garcia2020}. Several platforms follow model-driven engineering principles \cite{Hartsell2019}, for instance a model representation for production ML models \cite{Deak2018}. These types of platforms are relevant for the cyber-physical systems domain, for which we also found dedicated tool support \cite{Hartsell2019, Rill2019}.

With six papers, the third-largest category is formed by studies that elicited and described \textit{design challenges} for AI-based systems, e.g. by conducting empirical studies like surveys \cite{Zhang2020} or reporting industry experiences \cite{Haldar2019}. The scope of the described design challenges varies greatly and covers areas such as intelligent automotive systems \cite{Lan2018}, ML model management \cite{Schelter2018}, or ML fairness \cite{Holstein2019}. The details of these challenges are explained in Section 7 (RQ4).

We also found four papers on \textit{design patterns} for AI-based systems: an architecture pattern to improve the operational stability of ML systems \cite{Yokoyama2019}, patterns to improve the safety of systems with ML components \cite{Serban2019}, an architecture pattern to manage N-version ML models in safety critical systems \cite{Machida2019}, and finally more abstract solution patterns to address recurrent business analytics problems with ML \cite{Nalchigar2019}.

Finally, we found two studies about technical debt and the design stage, for which we created the category \textit{design debt} \cite{Foidl2019, Alahdab2019}, and resources for teaching software design for AI-based systems \cite{Kaestner2020}.

\begin{tcolorbox}
    Key takeaways for software design:
    \begin{itemize}
        \item Many design strategies to cope with specific quality attributes, e.g. with safety or reliability
        \item Several concrete AI infrastructure proposals, e.g. for sharing models as microservices
        \item However, at the system level, there are few proposals for patterns, design standards, or reference architectures
    \end{itemize}
\end{tcolorbox}

\subsection{Software construction (23 studies)}
A total of 23 studies were concerned with \textit{software construction}. Many of these studies either provided specialized \textit{tools} (six papers) or more holistic \textit{platforms} (four papers) to support the development of AI systems. Exemplary application contexts for tools were the deployment and serving of general prediction systems~\cite{Crankshaw2017}, lowering the barrier for using ML techniques~\cite{Patel2010, Bailis2017, Kery2018, Khalajzadeh2018}, or model-based development toolchains~\cite{Hartsell2019}. For platforms, the common goal was to provide a more comprehensive infrastructure to improve and accelerate the AI development workflow~\cite{Bailis2017, Schleier-Smith2015, Colomo2019}, sometimes in more specialized domains like safety-critical robotics~\cite{Desai2019}.

Furthermore, eight studies reported construction \textit{practices and guidelines} for AI systems based on diverse experiences, such as implementing ML components to detect and correct transaction errors in SAP~\cite{Rahman2019}, a systematic comparison of DL frameworks and platforms~\cite{Guo2019}, experiences from improving Airbnb search results with DL~\cite{Haldar2019}, experiences from 150 ML applications at Booking.com~\cite{Bernardi2019}, AI model criteria relevant for end users~\cite{Fiebrink2011}, automatic version control in notebooks~\cite{Kery2018}, or practices collected via practitioner surveys and/or interviews~\cite{Wan2019, Zhang2020}. Similarly, seven studies reported current \textit{challenges} in constructing AI-based systems, most of them focusing on DL. Challenges have been derived via StackOverflow questions~\cite{Zhang2019, Islam2019}, surveys~\cite{Zhang2020}, theoretical analyses of the AI development process~\cite{Lan2018, Xie2018b}, or case studies~\cite{Arpteg2018, Raj2019}.

Lastly, one paper was concerned with the software construction of AI systems in a \textit{teaching} and education context~\cite{Kaestner2020}: Kästner and Kang describe their \enquote{SE for AI-enabled systems} course material and infrastructure and share lessons learned from educating Master students in this field.

\begin{tcolorbox}
    Key takeaways for software construction:
    \begin{itemize}
        \item Many state of practice studies synthesized construction challenges and guidelines to address them.
        \item Several tools and platforms have been proposed to improve AI development activities, but their maturity, rationales for their selection, and level of adoption remain vague.
    \end{itemize}
\end{tcolorbox}

\subsection{Software testing (115 studies)}
We identified 115 studies focused on \textit{software testing}. During the analysis, we formed 11 subcategories (\textit{bugs \& debugging}, \textit{challenges}, \textit{explanations}, \textit{quality testing}, \textit{test architecture \& languages}, \textit{test case}, \textit{test case generation}, \textit{test case selection}, \textit{testing methods}, \textit{testing metrics}, and \textit{generic}), which we partially refined into sub-themes. For each of the 115 studies, we assigned one or more of these subcategories. In three cases, we assigned five different subcategories to a paper. The different subcategories are described below:

Four papers addressed \textit{bugs \& debugging}. Three of these papers conducted empirical studies examining bugs in ML projects~\cite{Sun2017, Zhang2018b, Islam2019b}, whereas one paper proposed a specialized debugger for ML models~\cite{Cai2016}.

A total of 12 papers studied the \textit{challenges} in software testing for AI. Nine of them discussed the challenges, issues, and needs in AI software testing based on the current state of the art, either for generic AI systems~\cite{Otero2015, Breck2017, Huang2018, Schelter2018, Gao2020} or focusing on the particular challenges for autonomous vehicles or other safety-critical systems~\cite{Koopman2016, Salay2017, Lan2018, Kuwajima2020}. Finally, three proposals identified the challenges for generic AI or ML systems through empirical methods like questionnaire surveys with practitioners~\cite{Holstein2019, Ishikawa2019, Zhang2020}.

Two papers addressed testing-related \textit{explanations} for ML systems~\cite{Kulesza2015, Nushi2018}. Due to the  difficulty to understand the results of ML systems in some scenarios, these papers provided a method for explaining how the ML system reached a particular result, including failures or how the tester addressed them to correct the ML system.

In 35 papers, the focus of the presented testing approach was aimed at improving very specific quality characteristics of the AI-based system under test (subcategory \textit{quality testing}), including safety (e.g.~\cite{Chen2018, Aniculaesei2019, Salay2018}), robustness (e.g.~\cite{Evtimov2017, Juez2017, Srisakaokul2018b}), security (e.g.~\cite{Du2019c, Sarathy2019, Bozic2018}), fairness~\cite{Holstein2019, Udeshi2018, Aggarwal2019} or others (e.g.~\cite{Chakravarty2010, Koren2019, Rubaiyat2018}. Safety was indeed the most addressed quality characteristic with 21 proposals, followed by robustness and security with seven and four papers respectively. 

In the subcategory \textit{test architecture \& languages}, we identified one paper proposing a new testing architecture~\cite{Nishi2018}, and one paper proposing a specific testing language~\cite{Majumdar2019}.

17 papers were assigned to the subcategory \textit{test case}. The type of test cases that these studies addressed were adversarial test cases in 14 occasions (e.g.~\cite{Dreossi2018, Evtimov2017, Gopinath2017}) and corner test cases in 3 occasions~\cite{Yang2018, Bolte2019, Weibin2019}. As opposed to the related categories about test case generation and selection, papers in this general category were very focused on conceptualizing different types of test cases.

Regarding \textit{test case generation}, 40 papers provided automatic means for the generation of test cases. Most of them augment existing test cases, deriving new tests from an original dataset (e.g.~\cite{Dreossi2019, Tian2018, Jia2020}). Some of these proposals generate these test cases randomly (e.g.~\cite{Zhao2018, Majumdar2019}), but others focus on attaining specific objectives when generating test cases, like generating corner case testing inputs (e.g.~\cite{Yang2018}), adversarial testing inputs (e.g.~\cite{Evtimov2017, Yaghoubi2019, Zhou2018}) or increase the coverage of the test suites (e.g.~\cite{Du2019, Ma2019}). Other approaches generate test cases with discriminatory inputs to uncover fairness violations~\cite{Udeshi2018},  or with specific inputs to uncover disagreements between variants of an AI/ML model~\cite{Xie2019c}. Approaches like~\cite{Wolschke2017} generate test suites avoiding too similar test cases to minimize the number of tests to execute. It is worth mentioning that some proposals are based on simulation-based test generation, for instance, to generate tests for autonomous vehicles in simulated environments (e.g.~\cite{Tuncali2018, Gambi2019b, Juez2017}).

A total of 12 proposals were categorized with \textit{test case selection}. Some approaches proposed test case selection techniques as a complementary activity to the test case generation (e.g.~\cite{Fremont2020, Weibin2019, Du2019c}). One approach proposed a technique to select test cases based on a metric of importance~\cite{Gerasimou2020}, whereas others proposed techniques to identify corner cases~\cite{Bolte2019}, adversarial examples~\cite{Wang2018c} or likely failure scenarios~\cite{Koren2019}. Finally, a few approaches proposed techniques for test input prioritization to select the most important ones and reduce the cost of labeling~\cite{Byun2018, Feng2019} or reduce the performance cost of training and testing huge amounts of data~\cite{Spieker2019}.

A total of 36 papers addressed \textit{testing methods} for AI systems, following different techniques such as combinatorial testing~\cite{Klueck2018, Ma2019, Majumdar2019}, concolic testing~\cite{Sun2019c, Sun2018, Sun2020}, fuzzing (e.g.~\cite{Dreossi2019, Xie2019, Odena2019}, metamorphic testing (e.g.~\cite{Nakajima2019, Zhang2019, Dwarakanath2018}, or others (e.g.~\cite{Chakarov2016,  Dreossi2018, Ma2018d}. From the different methods used, it is interesting to point out that the most popular one is metamorphic testing with 16 studies, followed by fuzzing and mutation testing with six and five studies, respectively. 

Moreover, 20 papers focused on the definition and/or exploration of \textit{testing metrics} to measure the testing quality. 14 out of 20 focused on test coverage metrics (e.g.~\cite{Bolte2019, Hauer2019, Sun2019b}, whereas the rest of metrics were reported only by one study each: diversity~\cite{Srisakaokul2018b}, importance~\cite{Gerasimou2020}, suspiciousness~\cite{Eniser2019}, probability of sufficiency~\cite{Chakarov2016}, and disagreement~\cite{Xie2019c}.

Finally, 14 papers were categorized as \textit{generic}, as they did not address or contribute to a specific testing theme.

\begin{tcolorbox}
    Key takeaways for software testing:
    \begin{itemize}
        \item The main focus in software testing for AI is test cases (55 unique studies), including the two specialized areas test case generation (40) and test case selection (12).
        \item The majority of papers related to testing methods use metamorphic testing (16 out of 36), followed by fuzzing (6) and mutation testing (5).
        \item The majority of papers related to testing metrics propose novel coverage criteria (14 out of 20).
    \end{itemize}
\end{tcolorbox}

\subsection{Software maintenance (6 studies)}
The small \textit{software maintenance} area only comprises six studies, which we group further into three categories. Two studies empirically analyze the nature and prediction impact of \textit{bugs} in AI software~\cite{Sun2017,Leotta2019}. Similarly, two studies are concerned with providing specialized approaches or tool support for the \textit{debugging} of ML software by focusing on explanatory debugging in interactive ML~\cite{Kulesza2015}, and proposing their Tensorflow debugger based on dataflow graphs~\cite{Cai2016}. The remaining two papers elicited and reported maintenance \textit{challenges}, namely~\cite{Schelter2018} as an Amazon experience report in the area of ML model management, and~\cite{Zhang2020} via a questionnaire survey with DL practitioners.

\begin{tcolorbox}
    Key takeaways for software maintenance:
    \begin{itemize}
        \item Hardly any studies on the topic. More research is needed, as there are a few open challenges (see next section)
        \item In addition to state of practice analyses, the focus was on bugs in and debugging of AI-based systems.
    \end{itemize}
\end{tcolorbox}

\subsection{Software engineering process (31 studies)}
Many of the studies mapped to the \textit{SE process} area deal with an \textit{AI/ML process} (13 studies). Out of those, a few discuss processes in specific application areas, such as recommendation systems~\cite{Liu2017}. The remaining ones address processes in general. Although the majority of papers focuses on processes for developing AI-based systems, some papers (e.g.~\cite{Simard2017}) address the topic of processes for creating new ML algorithms and tools. Four papers~\cite{Hill2016, Liu2017, Santhanam2019, Bosch2020} present an overview of current practices and challenges faced during AI system development in comparison to traditional software development. For example, Hill et al.~\cite{Hill2016} conclude from their interviews with developers of AI systems that they generally struggle to establish a repeatable AI development process. Other authors address identified challenges by proposing concrete solutions or a general research agenda for AI systems engineering~\cite{Bosch2020}. Specific approaches include the application of agile development principles to AI model and system development~\cite{Schleier-Smith2015}, the integration of development and runtime methods known from DevOps~\cite{Aniculaesei2018, Martinez-Fernandez2020} or the adaptation of acknowledged process standards such as ISO 26262~\cite{Salay2018}.

Closely related to the AI/ML process is the \textit{AI/ML pipeline} category (6 studies), which instead of the overall AI system development process addresses only the part devoted to creating AI models~\cite{VanDerWeide2017, Cheng2018, Amershi2019, Nascimento2019, Hummer2019, Garcia2020}. 

We identified three papers that propose \textit{frameworks} for end-to-end support of AI system development~\cite{Colomo2019, Moreb2020, Raji2020}. The proposed frameworks target different concepts, e.g. software development, ML, algorithms, and data. They also have been designed for different domains and contexts, such as ML-based health systems~\cite{Moreb2020}, accountability improvement via algorithmic auditing~\cite{Raji2020}, and the support of ML solution development within digitalization processes~\cite{Colomo2019}.

Furthermore, we identified six studies which provide \textit{tools} to support the SE process of AI-based systems. An example are Patel’s general-purpose tools to provide AI/ML developers with structure for common processes and pipelines~\cite{Patel2010}. More specific use cases are covered by an ML platform to support iterative and rapid development of ML models~\cite{Schleier-Smith2015}, the DEEP platform with a set of cloud-based services for the development and deployment of ML applications~\cite{Garcia2020}, and a toolbox to support data-driven engineering of neural networks for safety-critical domains~\cite{Cheng2018}. Lastly, other works envisioned how these platforms should be implemented~\cite{Bailis2017, Pedroza2019}.

There are also three studies which report \textit{best practices} to build ML-based systems. Mattos et al. propose five tactics to address challenges during the development of ML-based systems~\cite{Mattos2019}: minimum viable and explainable model; randomization; disabling imputation in early stages; automation after the prototype validation; and continuous experimentation. Additionally, experiences on large scale real-world ML-based systems from Microsoft~\cite{Amershi2019} and IBM~\cite{Akkiraju2018} have led to the proposal of \textit{maturity models}.

As in other SWEBOK areas, several studies talk about \textit{challenges} of AI development processes~\cite{Menzies2020, Lwakatare2019, Mattos2019, Santhanam2019, Liu2017, Kim2018, Flaounas2017} or a \textit{roadmap} to address them~\cite{Bailis2017}. The challenges are reported later in  Section~\ref{sec:results-rq4}.

\begin{tcolorbox}
    Key takeaways for SE process:
    \begin{itemize}
        \item Diverse researchers, including R\&D from large companies, have investigated the process to develop and maintain AI-based systems. Many of them highlight the need to form multidisciplinary teams for effective AI processes, e.g. including software engineers and data scientists.
        \item Many analyzed processes have been constructed in an ad-hoc manner based on the early experiences of large companies in AI-based systems.
        \item However, six studies have focused on AI pipelines at the model rather than the system level.
        \item Process-related support is emerging with tools (6 studies) and frameworks (3 studies).
    \end{itemize}
\end{tcolorbox}

\subsection{Software engineering models and methods (38 studies)}
From the 38 unique primary studies in this category, the majority formulate concrete proposals on models and methods, while a few also elaborate on challenges, practices, and roadmaps. In terms of topics, the papers lean a little more towards verification \& validation (V\&V) methods (24 papers) than models (12 papers, one of them also in the former V\&V methods category), with three papers reporting generic challenges, practices, and roadmaps. In general, there was a strong dominance of safety as a non-functional aspect and application domains like autonomous vehicles in this SWEBOK area.

Most of the papers in the \textit{V\&V methods} category (18 out of 24) focus on formally verifying safety of cyber-physical systems, such as autonomous vehicles or robotic systems containing AI components. Typically, the AI components are controlled by artificial neural networks (16 papers), in particular deep networks. Two papers~\cite{Ingrand2019, Seshia2016} present an overview of current challenges and practices in V\&V of autonomous systems, in particular those based on DL. Remaining papers propose alternative V\&V approaches to ensure correct, robust, and safe AI-based systems. Part of them aim at systems based on specific types of artificial networks, such as multi-layer feed-forward perceptron~\cite{Pulina2010} or recurrent neural networks~\cite{Du2019}; remaining papers provide generic approaches independent of specific ML methods being evaluated. One group of V\&V solutions aim at specific problems of networks being easily fooled by adversarial perturbations, i.e., minimal changes to correctly classified inputs, that cause the network to misclassify them~\cite{Huang2017}. These approaches explore space of adversarial counter-examples to identify and ensure safe regions of the input space, within which the network is robust against adversarial perturbations~\cite{Gopinath2017}. Solutions propose direct search for counter-examples or investigation of input space margins and corner cases. Counter-example or guaranteed ranges of inputs on which artificial neural networks correctly are searched using various optimization techniques. Novelty of these approaches lies often in how the optimization problem is formulated and solved. Computation complexity and scalability are typical problems faced in this area. More recent papers attempt to solve these issues, e.g.~\cite{Wang2018}.

In the \textit{models} category, we found primary studies targeting \textit{formal model analysis}, model-driven engineering, model reuse, and practices. Five papers focus on formal model analysis with different goals: analysis of safety and scalability in models for autonomous vehicles~\cite{Shalev-Shwartz2017}, quantitative analysis for systems based on recurrent neural networks~\cite{Du2019}, improvement and certification of robustness for ML models~\cite{Yang2020}, and approaches for formally checking safety~\cite{Wang2018} or security properties of neural networks~\cite{Wang2018b}. Furthermore, four papers study the use of models as the initial step for derivation of other artefacts (\textit{model-driven engineering}). Examples are the development of big data ML software~\cite{Koseler2019} and the incorporation of safe and robust control policies in ML models~\cite{Ghosh2019}. The other two from the same authors~\cite{Tuncali2018, Tuncali2019} target the derivation of testing frameworks for evaluating properties of an autonomous driving system with ML components. Lastly, we identified three primary studies addressing \textit{model reuse} using different approaches, such as an analysis of current model reuse practices and challenges for building systems based on artificial neural networks~\cite{Ghofrani2019}, and tools to retain and reuse implementations of deep neural networks~\cite{Sato2018}.

\begin{tcolorbox}
    Key takeaways for SE models and methods:
    \begin{itemize}
        \item 18 papers cover the verification and validation of cyber-physical systems with AI components.
        \item Safety as a non-functional aspect and application domains like autonomous vehicles are very prevalent in this area.
    \end{itemize}
\end{tcolorbox}

\subsection{Software quality (59 studies)}
\textit{Quality management} is a very broad area of SE, including a number of topics such as specifying quality requirements, measuring / assessing quality, and assuring quality (SWEBOK). So far, quality management during engineering of AI-based systems has been dominated by testing and formal verification methods (see sections on \textit{testing} and \textit{models and methods} above). Only 17 publications address the topic of defining and assessing quality of AI-based systems. Among these, two papers~\cite{Sculley2015, Foidl2019} discuss software technical debt, a derivative of software quality which refers most commonly to increased costs for maintenance and evolution due to earlier quality deficits. In particular, the authors warn software engineers tempted by quick wins of data-driven software systems of forgetting that these wins are not coming for free. To avoid incurring significant technical debt in terms of ongoing maintenance costs with AI systems, the authors explore technical debt-related risks, e.g., related to a software system itself as well as to associated data and data management systems.

Furthermore, five articles propose \textit{ML quality certification and assessment} to mitigate quality risks of deployed AI systems~\cite{Nakajima2019d, Yang2020, Jenn2020, Arnold2018, Raji2020}. Several of these investigate challenges of certifying AI systems and look for potential solutions in traditional safety-critical domains such as automotive, avionics, or railway, in particular how certification approaches in these domains evolved to adjust to technological advances. Proposed certification approaches include the assessment of development processes (including workflows and engineering choices) and their impact on the quality of delivered outcomes~\cite{Jenn2020}. The quality of AI systems is viewed from various perspectives, e.g., prediction performance quality, training mechanism quality, and lifecycle support quality including continuous operations~\cite{Nakajima2019d}. Inspired by declarations of conformity -– multi-dimensional fact sheets that capture and quantify various aspects of the product and its development to make it worthy of consumers' trust -– authors propose that AI service providers publish similar documents containing purpose, performance, safety, security, and provenance information for their customers~\cite{Arnold2018}.

The most commonly discussed quality characteristics include safety and related aspects such as robustness or explainability. In addition to specific quality characteristics, meta-characteristics of AI systems, such as provability (extent to which mathematical guarantees can be provided that some functional or non-functional properties are satisfied) or monitorability (extent to which a system provides information that allow to discriminate \enquote{correct} from \enquote{incorrect} behavior)~\cite{Jenn2020}, are discussed in this context as prerequisites for quality assessment and certifications. Several articles investigate quality aspects specifically relevant for AI-based systems, mostly based on important new challenges that software and requirements engineers must address when developing AI systems. Major trends in our sample are ML-specific quality aspects, such as \textit{ML safety}~\cite{Burton2017, Amodei2016, Shafaei2018, Hains2018, Saunders2017}, \textit{ML ethics}~\cite{Kumar2020, Bryson2017, Henderson2018, Coates2019}, and \textit{ML explainability}~\cite{Ribeiro2016, Jentzsch2019, Nushi2018}. Additionally, three articles from the same team of authors~\cite{Kuwajima2019c, Kuwajima2019b, Kuwajima2020} discuss how individual AI quality aspects relate to each other in the context of ISO 25000~\cite{ISO25010} as an established SE quality model. They also propose adaptations to the standard and how to quantitatively measure some AI quality aspects.

Due to the differences between AI-based systems and \enquote{traditional} software systems, six studies covered the \textit{update of the ISO 26262 standard} to address this. Contributions range from analyzing the deficiencies of the current version of ISO 26262~\cite{Salay2018, Salay2017, Koopman2016, Gharib2018}, to concrete adaptation proposals~\cite{Henriksson2018}, or a methodology framework for identifying functional deficiencies during system development~\cite{Chen2018}.

With 11 primary studies, \textit{ML quality assurance frameworks} constitute another important topic. These frameworks normally focus on specific quality aspects of ML products, such as allowability, achievability, robustness, avoidability and improvability~\cite{Nishi2018}, safety~\cite{McDermid2019, Desai2019}, specific safety issues like forward collision mitigation based on the ISO 22839 standard~\cite{Feth2017}, security~\cite{Du2019c}, algorithmic auditing~\cite{Raji2020}, robustness diversity~\cite{Srisakaokul2018b}, data validation~\cite{Breck2019}, or the reconciliation of product and service aspects~\cite{Nakajima2018}. Other approaches focus on continuous quality assurance with simulations~\cite{Arnold2018} and on run-time monitoring to manage identified risks~\cite{Koopman2018}. Furthermore, four primary studies explore assurance cases. Ishikawa et al. discuss the use of arguments or assurance cases for ML-based systems~\cite{Ishikawa2018}, including a framework for assessing the quality of ML components and systems~\cite{Ishikawa2018b}. Assurance cases have been also used in arguing the safety of highly automated driving functions, e.g. to solve underspecification with graphical structuring notation~\cite{Gauerhof2018} or to address functional insufficiencies in CNN-based perception functions~\cite{Burton2017}.

Four studies focus on \textit{ML defects}, i.e. several researchers have studied the specific types of bugs in AI-based systems~\cite{Islam2019, Thung2012, Chen2018, Leotta2019}. Similarly, five articles discuss \textit{ML attacks}, mostly with a focus on the use of adversarial examples~\cite{Dreossi2018, Evtimov2017, Ji2018, Srisakaokul2018b}, for instance by applying adversarial perturbations under different physical conditions in cyber-physical systems. Tramer et al. also discuss attacks to steal the complete models of AI-based systems~\cite{Tramer2016}.

As with other software systems, ML-based systems require \textit{monitoring}. We can find monitoring approaches combining ML with runtime monitoring to detect violations of system invariants in the actions' execution policies~\cite{Mallozzi2018}, managing identified risks, catching assumption violations, and unknown unknowns as they arise in deployed systems~\cite{Koopman2018},  and as a runtime safety~\cite{Feth2017} or ethical~\cite{Arnold2018} supervisors.

The remaining primary studies focus on either \textit{challenges} or establishing a research \textit{roadmap}, which is detailed in Section~\ref{sec:results-rq4}. We can highlight that in the 11 primary studies discussing roadmaps, they are often related to safety and the standard ISO 26262~\cite{Amodei2016, Hains2018, Koopman2016, Gharib2018, Varshney2017}. This seems to be a major challenge for which people not only work on detailed research contributions but see the need for a larger research roadmap. These roadmaps usually include suggestions for extensions of the standards and V\&V methods. There are two primary studies~\cite{Ji2018, Hains2018} that also address security and attacks. One roadmap combines it with the safety roadmap and calls for better integration of ML development into SE methods. The other roadmap concentrates on different types of attacks and countermeasures. Further explicitly mentioned quality attributes for which there is a roadmap are user experience~\cite{Yang2017} and fairness~\cite{Holstein2019}. One roadmap also discusses quality assurance certification~\cite{Arnold2018} and proposes to add FactSheets to ML services to increase trust. Finally, three primary studies~\cite{Sarathy2019, Kuwajima2019b, Varshney2017} propose roadmaps for general quality with ML-specific extensions to the ISO 25000 standard series. They include diverse aspects such as processes, V\&V methods, and formal analysis.

\begin{tcolorbox}
    Key takeaways for software quality:
    \begin{itemize}
        \item The specific quality aspects of AI-based systems have triggered the need to update standards such as ISO 25000 and ISO 26262.
        \item Most of the studies focus on ML quality attributes, frameworks, assurance and certification.
    \end{itemize}
\end{tcolorbox}

\subsection{Remaining SWEBOK areas (4 studies)}
Other SWEBOK areas are present to a lesser extent in our sample. In the \textit{SE management} area, Wolf et al. showed that AI software projects lead to dynamic and complex settings which necessitates active and engaged sensemaking~\cite{Wolf2020}: software teams must strive to create coherence between AI environments, AI model ecosystems, and the business contexts that emerge while building AI systems. In the second study in this area, Raji et al. introduced a framework for algorithmic auditing for end-to-end support during the internal AI system development life cycle~\cite{Raji2020}. For \textit{software configuration management}, we identified one paper: Schelter et al. presented experiences with ML model management at Amazon and outlined challenges~\cite{Schelter2018}. Furthermore, one paper was categorized as \textit{SE professional practice}. With the FactSheets approach from Arnold et al., AI service providers can publish documentation about their AI system including information on safety or data provenance to create an environment of transparency and trust~\cite{Arnold2018}. Lastly, we did not identify any primary study in the area of \textit{SE economics}, nor for \textit{computing}, \textit{mathematical}, and \textit{engineering foundations}.

\subsection{Discussion}
The aggregated results for RQ3 imply several notable findings, which we briefly discuss in this section.

\textbf{Observation 3.1: Many studies in our sample were related to software testing (115 / 248) and software quality (59 / 248).} These two SWEBOK areas received particular attention, implying that their state of research is much more advanced compared to the other areas. While testing- and quality-related challenges of AI-based systems are by no means completely solved, researchers active in these areas should be especially careful when selecting the scope of their contributions and positioning them regarding existing work. More focused literature studies can support such a fine-grained overview. While several such studies exist for the testing of AI-based systems (see Section 2.2), the diverse field of software quality in this area would still benefit from a detailed review, especially since most related studies in our sample are focused on approaches related to safety, robustness, or reliability in the context of cyber-physical systems like autonomous vehicles.

\textbf{Observation 3.2: The area of software maintenance is one of the smallest categories in our sample (6 / 248).} While a few studies from software quality may also touch maintainability, evolvability, or the concept of technical debt, software maintenance received very little attention overall. Considering the peculiarities of AI-based systems and the importance of an effective and efficient maintenance and evolution for any long-living production system, this may constitute an important research gap. A reason for this could be that most researchers and practitioners still focus on the effective initial creation of these systems and may not yet have considered optimizing their maintenance. Additionally, many AI-based systems of the new wave are not that old and may therefore not yet require sophisticated approaches necessary for, e.g. decades-old code bases.

\textbf{Observation 3.3: In the SE process area, we can find recent important contributions, but rather context-specific than widely adopted ones.} We believe that multidisciplinary research in this area is needed to integrate data collection and AI modeling into the SE lifecycle and vice versa. We did not find a standard process nor manifesto in the primary studies in this direction. Also, some processes mainly evolved from the data mining area (e.g., CRISP-DM), lacking an SE perspective. 

\textbf{Observation 3.4:  Most discussed SWEBOK areas include several recent state-of-practice studies to identify concrete peculiarities for the development of AI-based systems (e.g. via surveys, interviews, StackOverflow mining, etc.).} Researchers are still in the process of discovering and analyzing challenges and practices in this field, indicating that SE4AI research continues to be in a formative stage. Since finding out what techniques practitioners in this area actually use and what concrete problems they face is essential, there is still the need for additional studies like this, especially in less prevalent SWEBOK areas identified by us.

\textbf{Observation 3.5: The majority of identified studies are only concerned with a single SWEBOK area (189 / 248).} While there is much value to be gained from studies with such a detailed focus, the  creation of a successful AI-based system requires an effective interplay and connection between the majority of SWEBOK areas. While some studies from the SE process area also took such a perspective, we identified only very few holistic studies in total. As the SE4AI field matures, there may be much potential for approaches that incorporate the different SE facets for AI-based systems in their entirety.

\section{RQ4: What are the existing challenges associated with SE for AI-based systems?}\label{sec:results-rq4}

As detailed in Section 3, the 39 papers that we analyzed in RQ4 included 94 challenges. A challenge may be classified into more than one SWEBOK Topic, although most of the challenges (70\%) were classified under one Topic only. %, see Figure~\ref{fig:Figure_7_1}, left. In total, these 94 challenges were assigned into 130 instances of SWEBOK Topics. Figure~\ref{fig:Figure_7_1}, right, shows the number of such Topic assignments that emerge in every paper, which varies from only 1 in five papers to 19 Topic assignments in one paper.

%\begin{figure}
 %   \centering
  %  \begin{center}\fbox{Figure 7.1}\end{center}
    %\includegraphics[width=\textwidth]{rq4-swebok-challenges.pdf}
   % \caption{Number of SWEBOK Topics associated with a challenge (left) and number of assignments of challenges to SWEBOK Topics per paper (right).}
    %\label{fig:Figure_7_1}
%\end{figure}

Table~\ref{tab:Tab_7_2} summarizes the number of papers, challenges and assignments to every SWEBOK Knowledge Area. Some Knowledge Areas prevale, although it cannot be said that a Knowledge Area excels significantly from the rest; furthermore, the order is different if we focus on the papers or on the challenges, because as said above several papers identify a number of challenges related to one particular Knowledge Area.

\begin{table}
\caption{Distribution of challenges into SWEBOK Knowledge Areas.}
\begin{tabular}{|l|c|c|c|}
\hline
\textbf{Category} & \multicolumn{1}{l|}{\textbf{\#papers}} & \multicolumn{1}{l|}{\textbf{\#challenges}} & \multicolumn{1}{l|}{\textbf{\#assignments}} \\ \hline
Software Engineering Models and Methods & 11 & 17 & 17 \\ \hline
Software Quality & 9 & 14 & 15 \\ \hline
Software Construction & 9 & 9 & 10 \\ \hline
Software Testing & 8 & 14 & 19 \\ \hline
Software Design & 7 & 10 & 10 \\ \hline
Software Runtime Behaviour & 6 & 10 & 10 \\ \hline
Software Requirements & 5 & 16 & 18 \\ \hline
Software Engineering Process & 5 & 8 & 12 \\ \hline
Software Engineering Professional Practice & 4 & 11 & 12 \\ \hline
Software Maintenance & 3 & 3 & 3 \\ \hline
Software Engineering Economics & 2 & 3 & 3 \\ \hline
Software Configuration Management & 1 & 1 & 1 \\ \hline
\end{tabular}
\label{tab:Tab_7_2}
\end{table}

If we look at the SWEBOK Topics, we see that some of them are quite popular. Table~\ref{tab:Tab_7_3} shows those referenced by 4 challenges or more. Remarkably, three of the topics that we proposed as an extension of SWEBOK appear in the top 5 positions, which is somehow natural (they naturally emerged because they were popular).

%% THE FIGURE IS A SUBSTITUTE OF TABLE TAB_7_2 &&
\begin{figure}
    \centering
    \includegraphics[width=\textwidth]{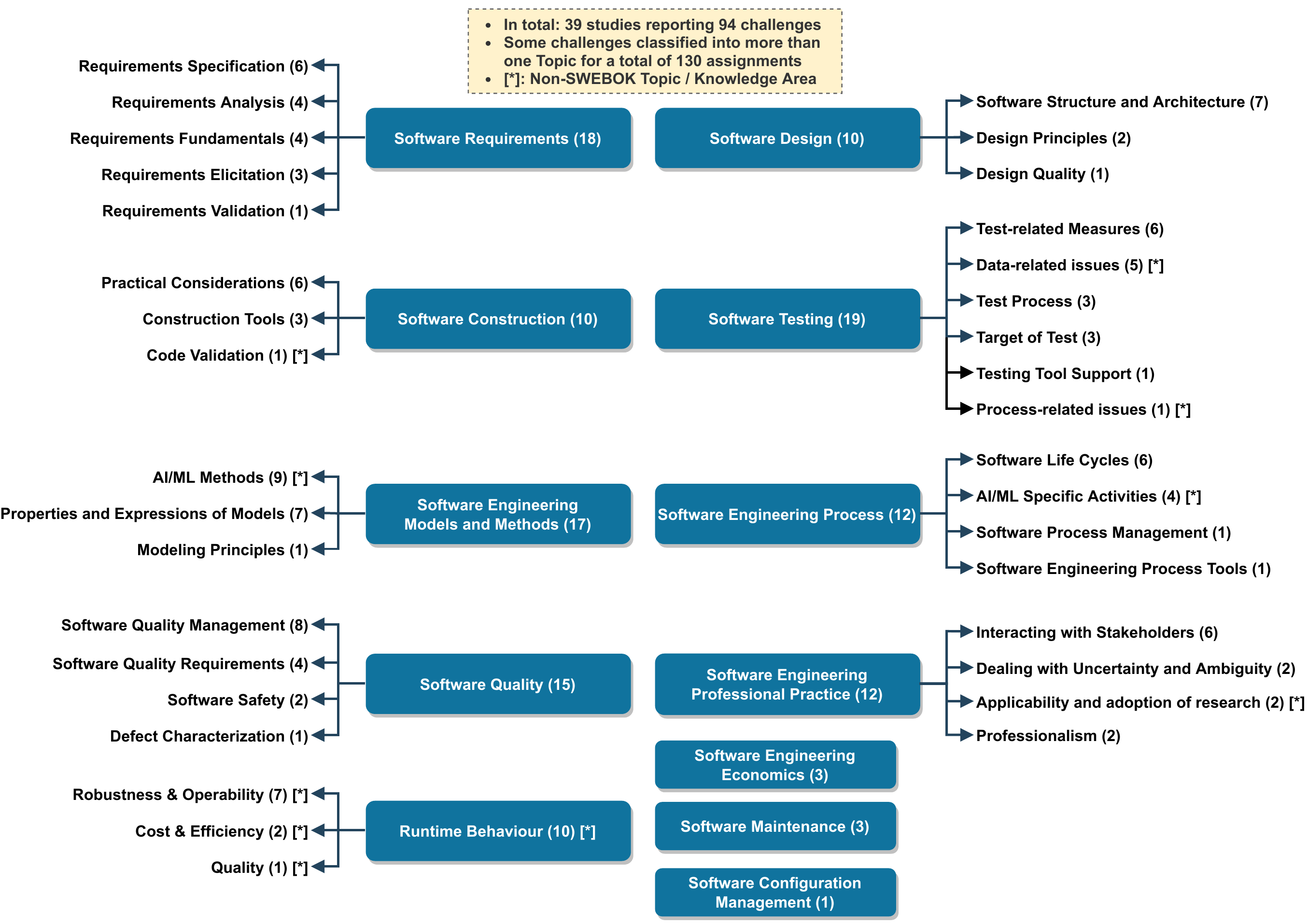}
    \caption{Distribution of Challenges into SWEBOK Knowledge Areas.}
    \label{fig:Figure_7_2}
\end{figure}

\begin{table}
\caption{Most popular SWEBOK Topics as referenced in the challenges.}
\begin{tabular}{llc}
\hline
\multicolumn{1}{|l|}{\textbf{SWEBOK Topic}} & \multicolumn{1}{l|}{\textbf{SWEBOK Knowledge Areas}} & \multicolumn{1}{c|}{\textbf{\#challenges}} \\ \hline
\multicolumn{1}{|l|}{AI/ML methods*} & \multicolumn{1}{l|}{Software Engineering Models and Methods} & \multicolumn{1}{c|}{9} \\ \hline
\multicolumn{1}{|l|}{Properties and expressions of models} & \multicolumn{1}{l|}{Software Engineering models and methods} & \multicolumn{1}{c|}{7} \\ \hline
\multicolumn{1}{|l|}{Robustness \& operability*} & \multicolumn{1}{l|}{Runtime Behaviour} & \multicolumn{1}{c|}{7} \\ \hline
\multicolumn{1}{|l|}{Interacting with stakeholders} & \multicolumn{1}{l|}{Software Engineering Professional Practice} & \multicolumn{1}{c|}{6} \\ \hline
\multicolumn{1}{|l|}{Data related issues*} & \multicolumn{1}{l|}{Software Testing} & \multicolumn{1}{c|}{5} \\ \hline
\multicolumn{1}{|l|}{Software quality assessment} & \multicolumn{1}{l|}{Software Quality} & \multicolumn{1}{c|}{5} \\ \hline
\multicolumn{1}{|l|}{Software structure and architecture} & \multicolumn{1}{l|}{Software Design} & \multicolumn{1}{c|}{5} \\ \hline
\multicolumn{1}{|l|}{AI/ML specific activities*} & \multicolumn{1}{l|}{Software Engineering Process} & \multicolumn{1}{c|}{4} \\ \hline
\multicolumn{1}{|l|}{Quality requirements} & \multicolumn{1}{l|}{Software Quality} & \multicolumn{1}{c|}{4} \\ \hline
\multicolumn{1}{|l|}{Requirements specification} & \multicolumn{1}{l|}{Software Requirements} & \multicolumn{1}{c|}{4} \\ \hline
\multicolumn{3}{l}{* Proposed extension of SWEBOK.}
\end{tabular}
\label{tab:Tab_7_3}
\end{table}

We describe below the challenges of each Knowledge Area\footnote{Names of SWEBOK Knowledge Areas and Topics appear in italics}. As for reporting style, we have opted to include the challenges respecting the words of the primary studies’ authors; therefore, our work has consisted mostly in grouping and articulating these challenges into a unifying narrative. This also implies that we are intentionally refraining from adding our own interpretations or enriching in any way the challenges identified in the primary studies. Last, for readability, we do neither quote the challenges nor include all citations in the text; Table~\ref{tab:Tab_7_4} includes the references for every SWEBOK Knowledge Area.

\begin{table}
\caption{Primary studies containing challenges per SWEBOK area.}
\begin{tabular}{|l|l|}
\hline
\textbf{SWEBOK Knowledge Area} & \textbf{List of references} \\ \hline
Software Engineering Models and Methods & \begin{tabular}[c]{@{}l@{}}\cite{Belani2019, Gao2020, Hains2018, Holstein2019, Ishikawa2019, Jenn2020, Koopman2016, Lan2018, Raj2019, Schelter2018, Shafaei2018}\end{tabular} \\ \hline
Software Requirements & \begin{tabular}[c]{@{}l@{}}\cite{Horkoff2019, Ishikawa2019, Jenn2020, Vogelsang2019, Zhang2020}\end{tabular} \\ \hline
Software Testing & \begin{tabular}[c]{@{}l@{}}\cite{Breck2017, Gao2020, Huang2018, Otero2015, Sarathy2019, Schelter2018, Shafaei2018, Zhang2020}\end{tabular} \\ \hline
Software Quality & \begin{tabular}[c]{@{}l@{}}\cite{Gao2020, Holstein2019, Horkoff2019, Ishikawa2019, Jenn2020, Kumar2020, McDermid2019, Sarathy2019, Shafaei2018}\end{tabular} \\ \hline
Software Engineering Professional Practice & \cite{Ingrand2019, Ishikawa2019, Sarathy2019, Vogelsang2019} \\ \hline
Software Construction & \begin{tabular}[c]{@{}l@{}}\cite{Horkoff2019, Ishikawa2019, Jenn2020, Kuwajima2019b, Lwakatare2019, Mattos2019, Menzies2020, Sarathy2019, Schelter2018, Zhang2020}\end{tabular} \\ \hline
Software Runtime Behaviour & \begin{tabular}[c]{@{}l@{}}\cite{Amodei2016, Horkoff2019, Lan2018, McDermid2019, Schelter2018, Shafaei2018}\end{tabular} \\ \hline
Software Engineering Process & \begin{tabular}[c]{@{}l@{}}\cite{Ishikawa2019, Kuwajima2019b, Liu2017, Lwakatare2019, Menzies2020}\end{tabular} \\ \hline
Software Design & \begin{tabular}[c]{@{}l@{}}\cite{Belani2019, Kumar2020, Liu2017, Lwakatare2019, Schelter2018, Zhang2020}\end{tabular} \\ \hline
Software Maintenance & \cite{Horkoff2019, Koopman2016, Schelter2018} \\ \hline
Software Engineering Economics & \cite{Ishikawa2019, Mattos2019} \\ \hline
Software Configuration Management & \cite{Schelter2018} \\ \hline
\end{tabular}
\label{tab:Tab_7_4}
\end{table}

\subsection{Software requirements}

Most of the challenges relate to one activity in the requirements engineering cycle, except a first subset related to fundamentals:

\begin{itemize}

\item {\textit{Fundamentals}. A number of challenges relate to functional and non-functional requirements fundamentals, arguing that: (i) our understanding of NFRs for ML is fragmented and incomplete, (ii) some new NFR types need to be considered, e.g. related to explainability and freedom for discrimination, (iii) measurement of functional requirements in practice for both functional and non-functional requirements is needed.}

\item {\textit{Elicitation}. The use of data as a source of requirements is attractive, but due to the high volume of such data, it requires tool support to detect features from massive data. Also, adopting the adequate stakeholder perspective is necessary, for instance to elicit explainability requirements, the user’s point of view needs to be adopted. Finally, it is crucial to elicit the characteristics (e.g., related to ethnicity or gender) that must not be used in AI-based systems to avoid discriminating samples.}

\item {\textit{Analysis}. The most important challenge is the need to negotiate upon unfeasible 100\% accuracy demands issued by customers. Moreover, it is mentioned that verifying the model features extracted from data is challenging . 
}

\item {\textit{Specification}. It is difficult to specify requirements for AI-based systems due to four main challenges: (i) transferring the problem definition into specifications, (ii) specifying clearly the concepts used by ML techniques (i.e., domain knowledge), (iii) understanding how requirements can be measured, (iv) making clear specifications from which testing and quality assurance activities can be derived. In the case of non-functional requirements, also the challenge of defining ML-specific trade-offs is mentioned.
}

\item {\textit{Validation}. This activity is endangered by the inherent uncertainty of the results produced by AI-based systems (e.g., relating to accuracy, cost, etc.), as well as by the difficulty to understand and use the notion of requirements coverage. Uncertainty of results has been said to be a source of anxiety for customers.
}

\end{itemize}

\subsection{Software design}

Software design challenges occur at different levels:

\begin{itemize}

\item {\textit{Design principles.} It is a challenge to overcome the CACE principle (Changing Anything Changes Everything), which is mainly due to the entanglement created by ML models.}

\item {\textit{Software structure and architecture.} Several situations particular to AI-based systems result in challenges to their structure: (i) ethics is a main issue of these applications, and a challenge is where to place the logic that governs ethical behaviour responding to criteria like scalability (needed with the advent of technologies such as 5G that demand huge amounts of sensors); (ii) AI-based systems need to deal with undeclared customers, who need to consume predictions of models; (iii) it becomes necessary to orchestrate different systems that are glued together in real-world deployments; (iv) it is requested to provide automatic exposure of ML metadata to automate and accelerate model lifecycle management; (v) it is needed to manage the consequences of the concurrent processing required by most AI-based systems, especially those implementing DL approaches. As a particular case in this Topic, the SWEBOK sub-Topic \textit{Design patterns} is also mentioned, given that the complexity of deploying ML techniques into production results in the emergence of several anti-patterns (glue code, pipeline jungles, etc.), making architecting a kind of plumbing rather than engineering activity.}

\item {\textit{Design Quality.} It is needed to reconcile conflicting forces, namely reproducibility, collaboration, ease of use and customization, in a single AI-based system.
}

\end{itemize}

\subsection{Software construction}

Contrary to the two former Knowledge Areas, challenges reported for software construction are of very diverse nature, most of them belonging to the \textit{Practical Considerations} Topic:

\begin{itemize}

\item {Implementation of ML algorithms involves several issues, among them strong dependency on data, high level of parallelism or use of complex tools like TensorFlow.}

\item {End-to-end AI-based systems often comprise components written in different programming languages, making the management of applications challenging (e.g., ensuring consistency with error-checking tools across different languages).}

\item {Implementation of AI-based systems with third-party components of any kind. This poses significant challenges in safety given that these components may not be assured with traditional methods, compromising their adoption in industries like avionics.}

\item {It is a challenge to control quality during the development of DL applications.}

\item {The Integration of AI/ML components with traditional software is also mentioned as a challenge from a quality perspective.}

\end{itemize}

A number of challenges are related to \textit{Software Construction Tools}:

\begin{itemize}

\item Companies need to struggle with non-flexible and functionally limited AI/ML development tools that are difficult to be incorporated within the company process, resulting in fragmented toolchains that hamper the work of business analysts, who can not generate ML prototypes quickly to experiment with.

\item The lack of tool infrastructure to support the development and deployment of DL solutions. This challenge is aggravated by the lack of expertise in usual IT teams to build this infrastructure.

\end{itemize}

Last, \textit{Code Validation} was mentioned as a challenge due to the difficulty in understanding and using the notion of requirements coverage when validating the code. 

\subsection{Software testing}

We found some challenges directly related to \textit{Key Issues} of software testing, majorly related to data and one also to process:

\begin{itemize}

\item Data and models cannot be strongly specified a priori, which means that testing of AI-based systems is dependent upon uncertain data and models.

\item Achieving scalability of testing with millions of parameters in AI/ML applications.

\item The way AI-based systems are developed, i.e. through a training phase, introduces the risk of overfitting training data.

\item Dealing with the inherent incompleteness of training and testing data is a major challenge that yields to insufficiencies when the context of execution is not fully represented in the training set. This is crucial in life-critical systems as autonomous vehicles.

\item Difficulty to collect enough data to test AI-based systems; for this reason, it is suggested to develop tools that augment real data while maintaining the same semantics.

\item Long-running experiments and complex interactions between pipelines of models makes traceability of results’ changes difficult to keep.

\end{itemize}

A number of challenges belong to the \textit{Test-Related Measures} Topic, with special emphasis on coverage:

\begin{itemize}

\item The need of systematic methods to prepare quality training and coverage-oriented datasets.

\item Understanding what coverage means and how to improve it are more focused challenges also reported in a couple of papers.

\end{itemize}

The rest of the challenges applies to the following topics:

\begin{itemize}

\item \textit{Test Process.} The generation of reliable test oracles and effective corner cases are process activities identified as challenging. As a practical consideration, it is reported that repeatability of test results is difficult to achieve due to the ability of AI-based systems to learn over time.

\item \textit{Target of Test.} It is reported that AI-based systems suffer from what is called the oracle problem: that ground truth data is difficult or sometimes impossible to get. Furthermore, the problem of having millions of parameters mentioned above also impacts on this subcategory due to the inherent variability behind these parameters.

\item \textit{Testing Tool Support.} How to develop automatic solutions to support testing of AI-based systems.

\end{itemize}

\subsection{Software engineering process}

Most of the challenges are related to the \textit{Software Life Cycle} Topic, as follows:

\begin{itemize}

\item \textit{Life cycle models.} It is necessary to have highly iterative models that allow to evolve solutions quickly. Going further, there is a need for continuous engineering because AI-based systems can be easily invalidated by trend changes. Development iterations are slowed down when re-assessing the models after introducing changes.

\item \textit{AI/ML specific activities.} This newly proposed Topic groups all challenges around activities that are specific to AI-based systems. In particular: (i) accurate and consistent annotation processes; (ii) ability to reproduce model selection experiments quickly; (iii) interleaving execution of experiments with interpretation of their results; (iv) exploration of different options of ML solutions.

\item \textit{Practical considerations.} First and foremost, AI-based software development requires software engineers. More focused challenges are: (i) the need of scaling models at the end of the life cycle (production); (ii) the difficulty of managing complex and poor logging mechanisms provided as part of the system infrastructure that makes analysis hard.

\end{itemize}

In addition, we find challenges related to:

\begin{itemize}

\item \textit{Software Process Management.} It is reported the challenge of managing complex workflows in charge of learning the ML models and bringing them into production.

\item \textit{SE Process Tools.} Linked to the challenge of annotation processes, it is a need to produce annotation tools for forming accurate and consistent annotations in large datasets.

\end{itemize}

\subsection{Software Engineering Models and Methods}

The \textit{Modeling} Topic has several challenges associated to its sub-topics:

\begin{itemize}

\item \textit{Modeling Principles.} It is necessary to avoid model overfit, which may be produced because of either: (i) training data having accidental correlations unrelated to the desired behaviour, or (ii) validation data not independent or diverse from the training data in every way except the desired features.

\item \textit{Properties and Expressions of Models.} First, data itself poses several challenges, as: (i) models need to rely upon high quality, properly curated datasets, as an indispensable requirement towards fairness of ML models; (ii) data should include as much as possible rare cases, which could entail learning problems, even considering that this rarity can make their collection expensive; (iii) data volumes and variety should be enough regarding the intended provided function. Second, the model built upon this data poses several difficulties: (i) it should capture behaviour that escapes the human eye (even at the cost of making validation more complex); (ii) related to this, the model provides results that are hard (if not impossible) to understand intuitively by humans; (iii) going further, the models are not just counter-intuitive but non-deterministic and providing uncertain predictions, and thus difficult to test. Last, the expression of the ML models is challenging due to the need of combining different models embedded therein) and also the difficulty of handling data dependencies when defining the models, which may convey different problems as instability and correction cascades, among others.

\item \textit{AI/ML methods.} We have identified this new topic that fits perfectly with the existing SE Methods’ subtopics in SWEBOK (namely, heuristic methods, formal methods, prototyping methods and agile methods). Challenges are: (i) expensiveness of data labeling; (ii) hard reasoning on robustness of ML techniques (especially worst case behaviour); (iii) development of systematic methods for preparing training and validation datasets; (iv) even in presence of these methods, ability to deal with incomplete training data; (v) model management in real-world ML deployments that involve complex orchestration frameworks or heterogeneous code bases written in different programming languages; (vi) automatic or semi-automatic formal verification of models; (vii) dealing with all aspects of data management, embracing data collection (e.g., lack of metadata), data exploration (e.g., heterogeneity of data sources), data preprocessing (e.g., cleaning of dirty data), dataset preparation (e.g., data dependencies), deployment (e.g., overfitting) and post deployment (e.g. feedback loops).

\end{itemize}

\subsection{Software Quality}

A first set of challenges are related to the \textit{Software Quality Management Process} in its three subtopics:

\begin{itemize}

\item \textit{Software Quality Assurance.} Challenges are manifold: (i) define quality assurance standards for AI-based systems and ensure that they scale well (e.g., they should adapt well to the continuous evolution of ML models and thus system behaviour); (ii) establish quality assurance criteria in the presence of big data (as required by AI-based systems); (iii) dealing with not assured components (e.g., third-party components or legacy software); (iv) assurance of safety and stability is particularly challenging because there are no clear principles established.

\item \textit{Verification \& Validation.} In general, verification and validation of the model produced by the training process regarding the intended function is challenging, principally because it needs to recognize the fact that the output of an ML model responding to a given input cannot be completely predicted.

\item \textit{Reviews and Audits.} Fairness auditing is threatened by the fact that it requires collecting information at individual-level demographics, which is rarely possible; new methods need  to adapt to this reality and allow demographics at coarser levels. 

\end{itemize}

A similar number of challenges appears concerning \textit{Practical Considerations of Software Quality Requirements}:

\begin{itemize}

\item Ensuring that the AI-based systems will not reinforce existing discriminations (on gender, race or religion).

\item Dealing with the limited knowledge of the effects that AI-based systems have on quality requirements (including their trade-offs).

\item Considering runtime quality when analysing the effects of AI-based systems on quality requirements.

\end{itemize}

An additional practical consideration emerges in the \textit{Defect Characterization} subtopic: the difficulty to explain to customers the failure in making a certain output due to the black-box nature of AI-based systems, especially when the real system output is counter-intuitive.

The last two challenges were specifically related to \textit{Software Safety}: (i) as stated above, absence of principles for quality of safety; (ii) oversimplification of safety assessment, not considering that it needs to remain valid through the application lifetime.

\subsection{Software Engineering Professional Practice}

We grouped the challenges related to professional practice into the following topics:

\begin{itemize}

\item \textit{Interacting with Stakeholders.} (i) We find challenges especially related with understanding of, and interaction with the customer because customers may have unrealistic expectations regarding the functionality, accuracy (requiring even 100\% accuracy) or adoption process of AI-based systems (expecting solutions starting to work with too little data available). At the end, it is necessary for the data scientists’ team to have the skills to interact with the customers and help them to set reasonable targets. (ii) From a more practical standpoint, due to the inherent iterative nature of AI-based systems that require continuous improvement of solutions, it becomes necessary to convince the customer about the need to keep paying continuously. (iii) Due to the blackbox nature of AI-based systems, it is a challenge to explain the customer failure to produce expected results.

\item\textit{ Applicability of Research in Practice.} We propose this new subcategory to group challenges related to transfer of research. Probably the most typical one (not only in the AI domain) is the oversimplification of reality when it comes to developing ML models. Nowadays, complex systems as autonomous systems have thousands of sensors and run several programs together, therefore toy academic examples (e.g., ``Lego Mindstorms'') are not acceptable in real settings. Also, it is mentioned the impediment of the risk perception of AI/ML results from the general public, which also needs to be considered when developing realistic AI-based systems.

\item \textit{Dealing with Uncertainty and Ambiguity.} Running ML/AI projects in real environments requires the ability to deal with the uncertainty of both estimating development time and cost, and validating the application considering that there is not a well-defined from any possible input to a given output.

\item \textit{Code of Ethics and Legal Issues.} Professional practice requires to consider these two aspects, which are challenging considering the intensive use of data by ML models (e.g., compliance to GDPR when dealing with personal data).

\end{itemize}

\subsection{Software Runtime Behaviour}

As explained in Section~\ref{sec:methodology} and \ref{sec:results-rq4}, we added this Knowledge Area that is not proposed in SWEBOK due to the importance of this aspect in AI-based systems. We further distinguished the following Topics:

\begin{itemize}

\item \textit{Robustness and Operability.} It refers to the fact that an AI-based system must be robust and easy to operate while in use. At this respect, mentioned challenges are: (i) the need to avoid negative side effects, so that the behaviour of the application does not damage its environment; (ii) in a similar vein, ensuring safe exploration during the learning process; (iii) preventing the hacking or gaming of reward functions, which could lead to artificially consider that the application reached its objectives when it is not true; (iv) adaptation of the application responding to changes in the operational environment; (v) dealing with unpredictable behaviour when the input of the application does not completely align with the training set (“distributional shift”); (vi) coordinate the workloads of the different systems that compose an ML pipeline at runtime.

\item \textit{Cost and Efficiency.} It tackles the problem of achieving efficient and cost-effective behaviour at runtime. We found two challenges: (i) provide scalable oversight for tasks for which there is insufficient information (by involving the human in the loop); (ii) overcome stringent timing and energy constraints which conflict with the resource-intensive nature of ML/AI applications.

\item \textit{Quality.} A challenge is to understand the effects of ML algorithms on desired qualities not only during ML solution design, but at runtime – during the lifetime of the ML solution.

\end{itemize}

\subsection{Remaining SWEBOK areas}

A handful of challenges related to some remaining SWEBOK Knowledge Areas:

\begin{itemize}

\item \textit{Software Maintenance.} All the challenges were related to training and validation data: (i) it is difficult to determine the frequency of retraining the models because training is usually conducted offline, therefore if there are changes in the context not captured by the training data, the model may become outdated and retraining is needed; (ii) as a follow-up of the preceding challenge, even minor changes on the training data may provoke a radical change in the learned rules, requiring thus complete revalidation of the model.

\item \textit{Software Configuration Management.} A main impediment to Software Building is the need of orchestrating different systems to deploy AI-based systems in a real context.

\item \textit{SE Economics.} Three challenges were reported relating to \textit{Risk and Uncertainty} and \textit{Value Management}: (i) how to quantify and assess the value of the knowledge generated in the company in the process of developing an AI-based system (even in the case that the application does not reach the initial goals); (ii) how to assess the long-term potential value of an AI/ML prototype, with the short-term metrics that can are usually gathered; (iii) how to manage the impossibility to make any prior guarantee on cost-effectiveness of AI-based systems due to uncertainty of their behaviour. 

\end{itemize}

\subsection{Discussion}

\textbf{Observation 4.1: Challenges are highly specific to the AI/ML domain.} Not only have we defined a non-SWEBOK Knowledge Area for covering runtime-related aspects, but also, we have classified in total 32 out of the 130 instances of challenges (i.e., 24.6\%) into non-SWEBOK Topics. This result provides evidence about the specificity of challenges reported by researchers and practitioners when it comes to AI-based systems and AI/ML model development. Another related observation is that artefacts that are widespread in the SE community as the SWEBOK body of knowledge are not totally fit to the AI-based systems engineering. Other artefacts which may suffer from similar drawbacks are quality standards as ISO 25010 and best practices as architectural or design patterns.

\textbf{Observation 4.2: Challenges are mainly of technical nature.} In general, more technical Knowledge Areas are those with more challenges identified, with the only exception of \textit{SE Professional Practice}, which in fact appeared as an accompanying Knowledge Area in five cases (in the same paper). Even in this Knowledge Area, the Topics identified were mainly two, namely Interacting with \textit{Stakeholders} and \textit{Dealing with Uncertainty and Ambiguity}, while other fundamental topics as \textit{Legal Issues}, \textit{Codes of Ethics} or \textit{Standards} were only accidentally mentioned or not mentioned at all. We may add the little importance given to \textit{SE Economics challenges}, with Topics as \textit{Risk}, \textit{Return on Investment} or \textit{Replacement and Retirement Decisions} not mentioned at all. Moreover, we did not identify challenges for the three foundations Knowledge Areas (computing, mathematical and engineering), although in this case, they may be hidden behind the technical challenges reported.

\textbf{Observation 4.3: Data-related issues are the most recurrent type of challenge.} Digging further into the technical nature of the challenges, we see a dominance of issues related to data, from the different perspectives provided by SWEBOK’s Knowledge Areas. We find challenges related to different stages of the software process (e.g., identifying features over a large amount of data during requirements elicitation, preparing high-quality training datasets during testing), and also to transversal activities as quality management (e.g., effects of data incompleteness on the overall system quality). In contrast, the surveyed papers have identified very few mitigation actions to cope with these challenges, e.g. generation and simulation of rare cases data to manage edge cases during learning. Therefore, the lack of such mitigation actions constitute research gaps to be addressed. This last observation aligns with other studies, e.g. Lwakatare et al. (2020) only presents 8 solutions for the 23 challenges that they identify in their paper.

\section{Threats to validity}\label{sec:threats}
As any other empirical study, ours faces a series of threats to validity. We report them below according to frequent threats specific to secondary studies \cite{ampatzoglou2019identifying} including mitigation actions. These specific threats are divided in three categories: study selection validity, data validity, and research validity. As a global mitigation action, we have considered in our study the ACM SIGSOFT Empirical Standards \cite{ralph2021empirical}, and in particular we have ensured: (i) to comply with all the eight essential specific attributes for systematic reviews; (ii) to avoid the three anti-patterns that apply to systematic reviews (i.e., not synthesising findings, not including quality assessment of primary studies, shortage of high-quality primary studies).

\subsection{Study selection validity}

Study selection threats can be identified in the steps 1 to 3 of our SMS, mainly conducting the search and screening of papers.

One of the inherent threats to any SMS is that it does not guarantee the inclusion of all the relevant works in the field. To mitigate this threat, we first conducted multiple pilot tests to build the search string to obtain the seed papers. Details of the different search strings options are available in our replication package. Four researchers independently undertook the preliminary search process before finalizing the search scope and search keywords. In a meeting with all researchers, initial search strings were discussed, and the reported one in Section 3 selected.

To mitigate sampling bias and publication bias, we used a combination of: (i) keyword automated searchers for seed papers and manual snowballing; (ii) check the annotated bibliography of an external prolific researcher; (iii) search in multiple indexes: Scopus for seed papers, and Google Scholar for snowballing considering pre-print servers (i.e., arXiv). We applied iteratively backward and forward snowballing until reaching 248 primary studies, which ensured that we covered a significant list of papers in a wide range of SE topics. As the field of SE4AI is further emerging and our resources for this study are limited, we decided to stop at this point.

By following the snowballing strategy, we also mitigated any possible threat related to the lack of a standard use of terminology, or lack of any relevant term in the search string, which is a common threat to validity in a pure string-search based methodology.

During the primary studies selection process, to mitigate any possible bias when applying the inclusion/exclusion criteria, these criteria were defined and updated in our protocol. Furthermore, each paper was assigned to two researchers from different institutions to decide about its inclusion or exclusion. Any disagreement was discussed between the two researchers, and if an agreement between the two was not reached, a third researcher was involved to make the final decision.

Regarding quality assessment, we used a classification of research types, including those without empirical evidence. Articles in this category would generally not be included in a systematic literature review, though in SMS they are important to spot trends of topics being worked on. We reported this in Section 4, and considered all papers as the outcome of an SMS is an inventory of papers on the topic area, mapped to a classification \cite{petersen2015guidelines}. Furthermore, we have also extracted information about the research rigor and industrial relevance of the primary studies.

\subsection{Data validity}
Data validity threats can be identified in the steps 4 and 5 of our SMS: keywording, and data extraction and mapping process.

During the process of data extraction, subjective bias may lead to the misclassification of data or an inconsistent interpretation of the extracted data by the researchers. To mitigate these risks, we piloted the data extraction form, conducted weekly meetings with all the researchers, and discussed potential issues related to data extraction. All found issues were discussed among all researchers and decisions were documented to ensure that all researchers followed consistent data extraction and synthesis criteria. Nonetheless, apart from the three studies used for piloting, each paper was extracted by a single researcher. While many extractions were fairly objective (e.g., a paper either explicitly described threats to validity or not, a paper used a specific term for AI-based system or a quality attribute goal, etc.), others left more room for interpretation. However, we argue that complete agreement is neither attainable for a sufficiently complex extraction process with multiple researchers nor is it strictly necessary, since we are very confident in the general tendencies and take-aways based on the extracted and synthesized data.

Furthermore, as explained in Section 3, we performed qualitative analysis through an existing conceptual framework (SWEBOK). We iterated on initial classifications among all researchers in our weekly meetings, leading to some proposals to update this framework.

Lastly, we need to mention that we adopted an inductive approach to the coding of properties. During the data extraction and mapping process, we e.g. extracted quality attribute goals and then grouped similar terms into unique codes. Including such terms explicitly in the search string may have produced slightly different results. Overall, we are confident that snowballing led to valid general tendencies in our sample, even though we do not claim completeness.

\subsection{Research validity}
Threats that can be identified in all steps of our SMS are classified as research validity.

We have performed a broad search of secondary studies (see Section 2.2). This allowed us to understand research gaps and brainstorm about the coverage and definition of our RQs. We have compared the results of our SMS to the current state-of-the-art within the scope of SE4AI, to generalize our findings with this scope.

To enable the reproducibility by other researchers of this SMS, we have documented all the steps performed, along with all the intermediate results. We have described the procedure in detail in Section \ref{sec:methodology}. Furthermore, all the raw materials and documented process are available in our replication package.

\section{Conclusions}

In this paper, we surveyed the literature for software engineering for artificial intelligence (SE4AI) in the context of the new wave of AI. In the last ten years, the number of papers published in the area of SE4AI has strongly increased. There were almost no papers up to 2015 while afterwards, we saw a strong increase to 102 in 2019. The share of more than 18\% on arXiv shows the “hotness” of the topic, but also emphasizes that literature reviews need to take arXiv into account. Furthermore, most articles are from a purely academic context, but 20\% of publications with only industry authors show the importance for practice. When we look at the countries of the authors, the United States play in a separate league altogether, while China, Germany, and Japan are the strongest of the remaining countries.

The empirical studies in our sample seem to form a healthy mix of case studies, experiments, and benchmarks. The latter play a larger role than in other fields of SE, which can be explained by the data-driven nature of the methods that often lend themselves to being benchmarked. In these studies, we see overall many realistic problems, data sets, and applications in practice. The involvement of practitioners improves some quality characteristics of the studies, like the realism of the case studies, and more significantly, their scale. We have also found, however, that authors often ignore discussing threats to validity. %This may be caused by the number of arXiv and workshop papers, which tend to discuss fewer threats to validity compared to journal or conference publications.

The terminology in the primary studies is all but homogeneous. This makes it often difficult to judge the scope of the contributions. We therefore propose to include a taxonomy in each SE4AI paper that clarifies the level of AI that the contribution is associated with. Furthermore, we suggest using the term \textit{AI component} if the article is about a part of a system that uses AI. An \textit{AI-based system} is a system consisting of various software and potentially other components with at least one AI component. %An \textit{AI library} is a library providing AI capabilities. 
Most of our primary studies are about AI-based systems or AI components. The most mentioned application domain for AI-based systems is automotive, while almost half of the contributions are not addressing any specific application domain. In terms of methods, almost all contributions use ML techniques, with DL as the largest explicitly mentioned technique.

Regarding the SE areas the primary studies contribute to, \textit{software testing} (115 studies) and \textit{software quality} (59 studies) are the most prevalent in our sample. Here, the main contributions are test cases for AI-based systems testing, and the need for update of current quality standards (e.g, ISO 26262) regarding AI-based systems. In a lesser extent, SE models and methods, SE processes, and software design have been investigated by more than 30 studies each. However, \textit{software construction}, \textit{software requirements}, and especially \textit{software maintenance} are less represented and seem to offer a lot of potential for research. Examples of studies in these areas are synthesizing best practices for AI-based systems construction, as well as holistic views for requirements engineering or debugging AI-based systems. Additionally, we identified several recent state-of-practice studies in every major SWEBOK area and very few holistic approaches spanning several areas, both of which seems to indicate that the SE4AI research field is still in a formative stage.

Challenges related to AI-based systems are mainly characterized by two facts. First, a significant share of identified challenges (25\%) are strongly tied to the system domain and thus difficult to classify using the SWEBOK Knowledge Areas and topics. Second, the mentioned challenges focus mostly on technical issues, especially data-related ones, and hardly consider challenges related to other areas such as economics.

We believe this is the most comprehensive survey of the SE4AI area for the new wave of AI. Yet, as the research field is still forming, updates to this comprehensive survey as well as more focused surveys for specific domains or SWEBOK areas will be needed. Moreover, for future work, we also want to survey the SE4AI field for earlier waves of AI to compare previous approaches with current research.

\begin{acks}
We thank Andreas Jedlitschka for his feedback on earlier stages of this work and for being supportive and useful to enhance it. This work has been partially funded by the "Beatriz Galindo" Spanish Program BEAGAL18/00064 and by the DOGO4ML Spanish research project (ref. PID2020-117191RB-I00). We are very grateful to our anonymous reviewers for their comments and suggestions.
\end{acks}

\bibliographystyle{ACM-Reference-Format}
\bibliography{references,referencesSection2}

\appendix

\section{Data availability}

The replication package contains the instruments used during the SMS: from the search string used to derive the start set of paper, to the data extraction form, and to the detailed classifications from the data analysis. It is available on \url{https://doi.org/10.6084/m9.figshare.14538324}.

\section{CRediT author statement}

\textbf{Silverio Martínez-Fernández}: Conceptualization, Methodology, Formal analysis, Investigation, Data Curation, Writing - Original Draft, Writing - Review \& Editing, Visualization, Project administration

\textbf{Justus Bogner}: Conceptualization, Formal analysis, Investigation, Data Curation, Writing - Original Draft, Writing - Review \& Editing, Visualization

\textbf{Xavier Franch}: Conceptualization, Formal analysis, Investigation, Data Curation, Writing - Original Draft, Writing - Review \& Editing, Visualization

\textbf{Marc Oriol}: Conceptualization, Formal analysis, Investigation, Data Curation, Writing - Original Draft, Writing - Review \& Editing, Visualization

\textbf{Julien Siebert}: Conceptualization, Formal analysis, Investigation, Data Curation, Writing - Original Draft, Writing - Review \& Editing, Visualization

\textbf{Adam Trendowicz}: Conceptualization, Formal analysis, Investigation, Data Curation, Writing - Original Draft, Writing - Review \& Editing, Visualization

\textbf{Anna Maria Vollmer}: Conceptualization, Formal analysis, Investigation, Data Curation, Writing - Original Draft, Writing - Review \& Editing, Visualization

\textbf{Stefan Wagner}: Conceptualization, Formal analysis, Investigation, Data Curation, Writing - Original Draft, Writing - Review \& Editing, Visualization

\end{document}